\documentclass[runningheads]{llncs}
\usepackage{amsmath}
\usepackage{amssymb}
\usepackage{wrapfig}
\usepackage{url}
\DeclareMathAlphabet{\mathrm}    {OT1}{cmr}{m}{n}
\DeclareMathAlphabet{\mathrmbf}  {OT1}{cmr}{bx}{n}
\DeclareMathAlphabet{\mathrmit}  {OT1}{cmr}{m}{it}
\DeclareMathAlphabet{\mathrmbfit}{OT1}{cmr}{bx}{it}
\DeclareMathAlphabet{\mathsf}    {OT1}{cmss}{m}{n}
\DeclareMathAlphabet{\mathsfbf}  {OT1}{cmss}{bx}{n}
\DeclareMathAlphabet{\mathsfit}  {OT1}{cmss}{m}{sl}
\DeclareMathAlphabet{\mathttbf}  {OT1}{cmtt}{bx}{n}
\newcommand{\keywords}[1]{\par\addvspace\baselineskip\noindent\enspace\ignorespaces{\bfseries Keywords:\,}#1}
\newenvironment{principle}[1]{\begin{flushleft}\normalsize\em{{\begin{bfseries}\begin{upshape} #1 Principle:\end{upshape}\end{bfseries}}\/}}
{\end{flushleft}}
\begin{document}
\title{System Consequence}
\author{Robert E. Kent}
\institute{Ontologos}
\maketitle
\begin{abstract}
This paper discusses system consequence, 
a central idea in the project to lift the theory of information flow to 
the abstract level of universal logic and the theory of institutions. 
The theory of information flow is a theory of distributed logic. 
The theory of institutions is abstract model theory. 
A system is a collection of interconnected parts, 
where the whole may have properties that cannot be known from an analysis of the constituent parts in isolation. 
In an information system, 
the parts represent information resources and 
the interconnections represent constraints between the parts. 
System consequence, 
which is the extension of the consequence operator from theories to systems, 
models the available regularities represented by an information system as a whole. 
System consequence (without part-to-part constraints) 
is defined for a specific logical system (institution) in the theory of information flow. 
This paper generalizes the idea of system consequence to arbitrary logical systems. 
\end{abstract}

\section{Introduction}\label{introduction}

We study the information flow of ontologies and related information resources by using the theory of institutions,
which provides an axiomatization of the notion of logical system.
The theory of information flow is centered on a particular logical system denoted {\ttfamily IF}. 
Institutions are based on Tarski's idea, in his semantic definition of truth, 
that the notion of satisfaction is central (Goguen and Burstall \cite{goguen:burstall:92}).
Ontologies are of two types: populated and unpopulated. 
Unpopulated ontologies contain theoretical information only, whereas
populated ontologies also contain semantic information. 
Semantic information is related to theoretical information through satisfaction.
At the most elemental level,
we represent theoretical information as types (universals), 
semantic information as instances (tokens, particulars), and
satisfaction as classification
``It is particulars, things in the world, that carry information; 
  the information they carry is in the form of types'' 
(Barwise and Seligman \cite{barwise:seligman:97}). 

Abstraction is used in the theory of institutions:
the details of the notions of 
formal language, sentence, semantic structure and satisfaction
are abstracted away
from their meaning in specific institutions such as first order logic:
languages, sentences and structures are atomic (elemental) notions 
and satisfaction is a composite notion that relates sentences to structures.
Abstraction is used in the theory of information flow:
types and tokens (instances) are atomic notions 
and classification is a composite notion that relates types to tokens.
This paper combines these uses of abstraction.

Notions of information flow are used in both theories.
The theory of information flow defines the invariance of classification under parameterized atomic flow,
the ``adjoint connection'' between token and type flow.
The theory of institutions defines the invariance of satisfaction under parameterized atomic flow,
the ``adjoint connection'' between structure and sentence flow.
The theory of information flow 
defines direct and inverse molecular flow of {\ttfamily IF} theories and {\ttfamily IF} (local) logics,
but does not exploit the ``adjoint connection'' between these.
The theory of institutions 
defines direct and inverse molecular flow of specifications,
and exploits the ``adjoint connection'' between these.
This paper extends these uses of information flow.

Until this paper, 
the theory of institutions has not combined semantics with formalism into something like a local logic.
Here, we define and discuss the notion of system consequence, 
as part of the effort to generalize the theory of information flow to the level of logical systems.
At the atomic level the paper generalizes
from type sets, classifications and sequents
to languages, structures and sentences, respectively.
At the molecular level it generalizes
from {\ttfamily IF} theories and {\ttfamily IF} (local) logics
to specifications and logics (generic, sound or composite), respectively.
Logics are 
unsound and/or incomplete 
semantic representions for various information resources such as
universal algebras, libraries, knowledge bases (data collections with formal description),
or physical (chemical) theories for a portion of the physical (chemical) world.

The paper falls into two parts:
section~\ref{sec:logical:systems} is concerned with the theory of institutions and
section~\ref{sec:channel:theory} is about channel theory.
Readers are assumed to be familiar with the basic notions of 
category theory as presented in Barr and Wells \cite{barr:wells:99} and
information flow as presented in Barwise and Seligman \cite{barwise:seligman:97}.
%
Section~\ref{sec:logical:systems} describes the two alternate representations for logical systems (institutions):
the heterogeneous representation is outlined in subsection~\ref{heterogeneous:representation} and
the homogeneous representation is discussed in some detail in subsection~\ref{homogeneous:representation}.
Important concepts of the heterogeneous representation are languages, sentences and structures.
Sentences are the atoms of formalism, and structures are the atoms of semantics.
Important concepts of the homogeneous representation are specifications and logics.
Specifications, the molecules of formalism, are partitioned into complete fiber preorders over their languages.
Logics, the molecules of semantics, are partitioned into complete fiber preorders over their structures.
Section~\ref{sec:logical:systems} discusses
the information flow between fibers of specifications along language morphisms 
                 and between fibers of logics along structure morphisms.
In addition,
this section also describes several well-known logical systems.
%
Section~\ref{sec:channel:theory} generalizes channel theory, the theory of information flow, to the theory of institutions.
Subsection~\ref{information:systems} defines distributed and information systems 
from both the formal and semantic perspectives.
Subsection~\ref{information:flow}
defines information channels over distributed systems and
describes direct and inverse information flow along channels.
System consequence is defined in terms of these notions of information flow.

\section{Logical Systems}\label{sec:logical:systems}

Any ontology is based on the logical {\em language} $\Sigma$ of a domain (of discourse), 
which often consists of 
the generic ideas of the connectives and quantifiers from logic and 
the specific ideas of the signature (the constant, function and relation symbols) for that domain. 
In the institutional approach,
a {\em sentence} is the atom of formalism and a {\em structure} is the atom of semantics.
Both sentence and structure are described and constrained by the logical language $\Sigma$. 
The collection of sentences and the category\footnote{A category $\mathrmbf{C}$
represents some ``species of mathematical structure'' (Goguen \cite{goguen:91}).
It consists of 
a collection of objects $|\mathrmbf{C}|$ which have that structure and 
a collection of morphisms, each directed from a source object to a target object, which preserve that structure
(there is an implicit notion of flow here).
Morphisms compose associatively,
and each object has an identity morphism on itself.
As examples,
$\mathrmbf{Set}$ is the category with sets as objects and functions as morphisms, and
$\mathrmbf{Cat}$ is the category with categories as objects and functors as morphisms.}
of structures 
are symbolized by $\mathrmbfit{sen}(\Sigma)$ and $\mathrmbfit{struc}(\Sigma)$, respectively.
A structure $M \in \mathrmbfit{struc}(\Sigma)$ 
provides a universe of discourse in which to interpret a sentence $s \in \mathrmbfit{sen}(\Sigma)$. 
In the context supplied or indexed by the language,
satisfaction is the composite connecting formalism and semantics.
A structure $M$ satisfies (is a model of) a sentence $s$ in the context of $\Sigma$,
symbolized $M \models_{\Sigma} s$ (a kind of triadic construct), when 
$s$ (holds in) is true when interpreted in $M$. 

In order to define the flow of information,
we make several assumptions.
We assume that information resides at a (possibly abstract) location;
such a location is represented by, or indexed as, a language $\Sigma$.
We assume that any two locations can be connected by a link;
such a location link is represented by or indexed as a language morphism
$\sigma : \Sigma_{1} \rightarrow \Sigma_{2}$, 
which has source language $\Sigma_{1}$ and target language $\Sigma_{2}$.
This is also a primitive notion in this paper.
The languages $\Sigma_{1}$ and $\Sigma_{2}$ represent two locations and 
the language morphism $\sigma$ enables information flow from $\Sigma_{1}$ to $\Sigma_{2}$. 
We assume that languages form the object collection 
(and their morphisms form the morphism collection)
of a language category 
$\mathrmbf{Lang}$ (Fig.~\ref{logical:system}).
Starting from this base,
we describe two equivalent representations for the notion of a {\em logical system} or {\em institution},
a heterogeneous representation and a homogeneous representation.

\subsection{Heterogeneous Representation}\label{heterogeneous:representation}

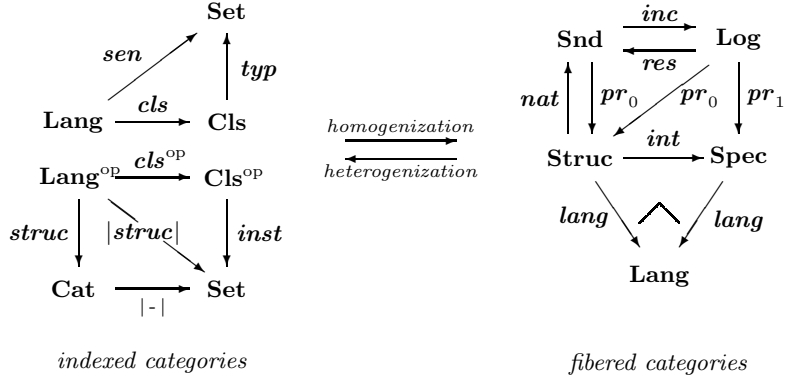
\begin{figure}
\begin{center}
\begin{tabular}[b]{c@{\hspace{60pt}}c@{\hspace{60pt}}c@{\hspace{5pt}}}
\\ 
\setlength{\unitlength}{0.7pt}
\begin{picture}(80,120)(0,0)
\put(0,75){\begin{picture}(80,60)(0,0)
\put(80,60){\makebox(0,0){$\mathrmbf{Set}$}}
\put(-3,0){\makebox(0,0){$\mathrmbf{Lang}$}}
\put(80,0){\makebox(0,0){$\mathrmbf{Cls}$}}
\put(40,10){\makebox(0,0){\footnotesize{$\mathrmbfit{cls}$}}}
\put(35,36){\makebox(0,0)[r]{\footnotesize{$\mathrmbfit{sen}$}}}
\put(87,28){\makebox(0,0)[l]{\footnotesize{$\mathrmbfit{typ}$}}}
\put(20,0){\vector(1,0){40}}
\put(80,12){\vector(0,1){36}}
\put(16,12){\vector(4,3){48}}
\end{picture}}
\put(0,-15){\begin{picture}(80,60)(0,0)
\put(1,60){\makebox(0,0){$\mathrmbf{Lang}^{\mathrm{op}}$}}
\put(84,60){\makebox(0,0){$\mathrmbf{Cls}^{\mathrm{op}}$}}
\put(-3,0){\makebox(0,0){$\mathrmbf{Cat}$}}
\put(80,0){\makebox(0,0){$\mathrmbf{Set}$}}
\put(-5,30){\makebox(0,0)[r]{\footnotesize{$\mathrmbfit{struc}$}}}
\put(44,70){\makebox(0,0){\footnotesize{$\mathrmbfit{cls}^{\mathrm{op}}$}}}
\put(85,30){\makebox(0,0)[l]{\footnotesize{$\mathrmbfit{inst}$}}}
\put(55,30){\makebox(0,0)[r]{\footnotesize{$|\mathrmbfit{struc}|$}}}
\put(40,-10){\makebox(0,0){\scriptsize{$|\,\mbox{-}\,|$}}}
\put(16,48){\line(4,-3){16}}
\put(46,25.5){\vector(4,-3){22}}
\put(0,48){\vector(0,-1){36}}
\put(80,48){\vector(0,-1){36}}
\put(20,60){\vector(1,0){40}}
\put(20,0){\vector(1,0){40}}
\end{picture}}
\put(40,-55){\makebox(0,0){\footnotesize{\emph{indexed categories}}}}
\end{picture}
&
\setlength{\unitlength}{0.7pt}
\begin{picture}(0,0)(-4,-60)
\put(0,12){\makebox(0,0){\scriptsize{\emph{homogenization}}}}
\put(-30,5){\vector(1,0){60}}
\put(30,-5){\vector(-1,0){60}}
\put(0,-12){\makebox(0,0){\scriptsize{\emph{heterogenization}}}}
\end{picture}
&
\setlength{\unitlength}{0.75pt}
\begin{picture}(80,120)(-10,8)
\put(0,120){\makebox(0,0){$\mathrmbf{Snd}$}}
\put(80,120){\makebox(0,0){$\mathrmbf{Log}$}}
\put(0,60){\makebox(0,0){$\mathrmbf{Struc}$}}
\put(40,0){\makebox(0,0){$\mathrmbf{Lang}$}}
\put(80,60){\makebox(0,0){$\mathrmbf{Spec}$}}
\put(15,30){\makebox(0,0)[r]{\footnotesize{$\mathrmbfit{lang}$}}}
\put(42,70){\makebox(0,0){\footnotesize{$\mathrmbfit{int}$}}}
\put(67,28){\makebox(0,0)[l]{\footnotesize{$\mathrmbfit{lang}$}}}
\put(-10,90){\makebox(0,0)[r]{\footnotesize{$\mathrmbfit{nat}$}}}
\put(10,90){\makebox(0,0)[l]{\footnotesize{$\mathrmbfit{pr}_0$}}}
\put(50,90){\makebox(0,0)[l]{\footnotesize{$\mathrmbfit{pr}_0$}}}
\put(84,90){\makebox(0,0)[l]{\footnotesize{$\mathrmbfit{pr}_{1}$}}}
\put(40,133){\makebox(0,0){\footnotesize{$\mathrmbfit{inc}$}}}
\put(40,107){\makebox(0,0){\footnotesize{$\mathrmbfit{res}$}}}
\qbezier(40,37)(45,32)(50,27)
\qbezier(40,37)(35,32)(30,27)
\put(-6,72){\vector(0,1){36}}
\put(6,108){\vector(0,-1){36}}
\put(80,108){\vector(0,-1){36}}
\put(22,126){\vector(1,0){36}}
\put(58,114){\vector(-1,0){36}}
\put(72,48){\vector(-2,-3){22}}
\put(22,60){\vector(1,0){40}}
\put(8,48){\vector(2,-3){22}}
\put(65,107){\vector(-4,-3){48}}
\put(40,-44){\makebox(0,0){\footnotesize{\emph{fibered categories}}}}
\end{picture}
\\ && \\ && \\ && 
\end{tabular}
\end{center}
\caption{Logical System}
\label{logical:system}
\end{figure}

The formal atoms (sentences) and the semantical atoms (structures) can be moved along language morphisms.
For any language morphism $\sigma : \Sigma_{1} \rightarrow \Sigma_{2}$, 
there is a sentence function 
$\mathrmbfit{sen}(\sigma) : \mathrmbfit{sen}(\Sigma_{1}) \rightarrow \mathrmbfit{sen}(\Sigma_{2})$
from the collection of source sentences to the collection of target sentences, and
there is a structure functor\footnote{A functor 
$\mathrmbfit{F} : \mathrmbf{A} \rightarrow \mathrmbf{B}$ is a 
``natural construction on structures of one species, yielding structures of another species'' 
(Goguen \cite{goguen:91}).
It is a link 
from a source category $\mathrmbf{A}$ of one species 
  to a target category $\mathrmbf{B}$ of another species,
which maps the source objects (morphisms) to target objects (morphisms),
preserving directionality, composition and identity.
As an example,
the underlying set functor 
$|\mbox{-}| : \mathrmbf{Cat} \rightarrow \mathrmbf{Set}$
maps a category to its collection of objects and 
maps a functor to its underlying function on objects.
The composition 
$\mathrmbfit{F} \circ \mathrmbfit{G} : \mathrmbf{A} \rightarrow \mathrmbf{C}$ 
of two functors
$\mathrmbfit{F} : \mathrmbf{A} \rightarrow \mathrmbf{B}$ and 
$\mathrmbfit{G} : \mathrmbf{B} \rightarrow \mathrmbf{C}$ 
is defined in terms of their object/morphism maps.}
$\mathrmbfit{struc}(\sigma) : \mathrmbfit{struc}(\Sigma_{2}) \rightarrow \mathrmbfit{struc}(\Sigma_{1})$
(in the contra direction)
from the category of target structures to the category of source structures.  
Hence, there is a sentence functor
$\mathrmbfit{sen} : \mathrmbf{Lang} \rightarrow \mathrmbf{Set}$
and a structure indexed category\footnote{
An indexed category $\mathrmbfit{C} : \mathrmbf{B}^{\mathrm{op}} \rightarrow \mathrmbf{Cat}$
is a (contravariant) functor from an indexing category $\mathrmbf{B}$ to $\mathrmbf{Cat}$.}
$\mathrmbfit{struc} : \mathrmbf{Lang}^{\mathrm{op}} \rightarrow \mathrmbf{Cat}$.
Passage composition with the underlying set functor yields the structure functor
$|\mathrmbfit{struc}| = \mathrmbfit{struc} \circ |\mbox{-}| 
: \mathrmbf{Lang}^{\mathrm{op}} \rightarrow \mathrmbf{Cat} \rightarrow \mathrmbf{Set}$ (Fig.~\ref{logical:system}).
The satisfaction relation is preserved during this information flow:
$\mathrmbfit{struc}(\sigma)(M_{2}) \models_{\Sigma_{1}} s_{1}$
\underline{iff}
$M_{2} \models_{\Sigma_{2}} \mathrmbfit{sen}(\sigma)(s_{1})$.
Equivalently,
using structure intent (Sec.~\ref{structures}),
$\mathrmbfit{struc}(\sigma)(M_{2})^{\Sigma_{1}} = \mathrmbfit{sen}(\sigma)^{-1}(M_{2}^{\Sigma_{2}})$.
In the institutional approach,
this is regarded as the invariance of truth under change of notation.
The formal and semantic functors, 
$\mathrmbfit{sen} : \mathrmbf{Lang} \rightarrow \mathrmbf{Set}$ and
$|\mathrmbfit{struc}| : \mathrmbf{Lang}^{\mathrm{op}} \rightarrow \mathrmbf{Set}$,
can be combined with satisfaction into a classification functor 
$\mathrmbfit{cls} : \mathrmbf{Lang} \rightarrow \mathrmbf{Cls}$,
where a language $\Sigma$ maps to the satisfaction classification 
$\mathrmbfit{cls}(\Sigma) 
= \langle |\mathrmbfit{struc}(\Sigma)|, \mathrmbfit{sen}(\Sigma), \models_{\Sigma} \rangle$
and invariance of truth under change of notation corresponds to the infomorphism condition.\footnote{
$\mathrmbf{Cls}$, 
the basic category of 
the theories of information flow (Barwise Seligman \cite{barwise:seligman:97}) and
formal concept analysis (Ganter and Wille \cite{ganter:wille:99}),
has classifications as objects and infomorphisms as morphisms.  
A classification $A = \langle{X,Y,\models}\rangle$
consists of a set of instances $X$, a set of types $Y$ 
and a binary incidence or classification relation $\models$ between instances and types.
An infomorphism
$f : A_{1} = \langle{X_{1},Y_{1},\models_{1}}\rangle
\rightleftharpoons \langle{X_{2},Y_{2},\models_{2}}\rangle = A_{2}$
consists of an instance function (in the contra direction)
$\check{f} : X_{2} \rightarrow X_{1}$
and a type function
$\hat{f} : Y_{1} \rightarrow Y_{2}$
that satisfy the condition
$\check{f}(x_{2}) \models_{1} y_{1}$ \underline{iff} $x_{2} \models_{2} \hat{f}(y_{1})$
for any source type $x_{1}$ and target instance $x_{2}$.}
The heterogeneous representation of logical systems 
is represented on the left side of Fig.~\ref{logical:system}.

\subsection{Homogeneous Representation}\label{homogeneous:representation}

This description is a way for ``homogeneously handling situations of structural heterogeneity'' 
(Goguen \cite{goguen:06}). 
It maps the heterogeneous situations represented by indexed categories 
to the homogeneous situations represented by fibered categories\footnote{A fibered category (fibration) 
$\mathrmbfit{P} : \mathrmbf{E} \rightarrow \mathrmbf{B}$
is a category $\mathrmbf{E}$ whose objects exist above some underlying base category $\mathrmbf{B}$. 
Objects $X$ in $\mathrmbf{B}$ index 
subcategories (often preorders) $\mathrmbfit{fbr}(X) \subseteq \mathrmbf{E}$ of the fibered category called fibers.
Links $f: X \rightarrow Y$ in $\mathrmbf{B}$ 
index contravariant inverse image pseudofunctors between fibers 
$\mathrmbfit{fbr}(f) : \mathrmbfit{fbr}(Y) \rightarrow \mathrmbfit{fbr}(X)$
taking fiber objects indexed by $Y$ to fiber objects indexed by $X$.
Pseudo means preservation of composition and identity up to natural isomorphism.}.
We describe fibered categories for structures, specifications and logics (generic or sound).
Specifications are the formal molecules of information that flow along language morphisms. 
Logics are the semantic molecules of information that flow along structure morphisms.
The fibered categories for specifications and logics 
are defined in terms of this information flow over 
the base categories of languages and structures,
respectively.
In all of the fibered categories described here,
the base category is ultimately the category of languages $\mathrmbf{Lang}$.
The homogeneous representation of logical systems (institutions) 
is represented on the right side of Fig.~\ref{logical:system}.

\subsubsection{Specifications.}\label{specifications}

An unpopulated ontology expressed in terms of a language $\Sigma$
is represented as a $\Sigma$-specification ($\Sigma$-presentation)
$T \in \mathrmbfit{spec}(\Sigma) = {\wp}\mathrmbfit{sen}(\Sigma)$~\footnote{The symbol `${\wp}$' 
denotes powerset for sets and direct image for functions.} 
consisting of a collection of $\Sigma$-sentences.
In the institutional approach,
a specification is a molecule of formalism,
which allows for the expression of the laws and facts deemed relevant for a domain. 
A structure $M	\in \mathrmbfit{struc}(\Sigma)$ satisfies (is a model of) a specification $T \in \mathrmbfit{spec}(\Sigma)$,
symbolized $M \models_{\Sigma} T$, when 
it satisfies every sentence in the specification,
$s \in T$ implies $M \models_{\Sigma} s$.
A specification $T$ entails a sentence $s$,
symbolized $T \vdash_{\Sigma} s$,
when any model of the specification satisfies the sentence.
The collection 
$T^{\scriptstyle\bullet} = \{s \in \mathrmbfit{sen}(\Sigma) \mid T \vdash_{\Sigma} s \}$
of all sentences entailed by a specification $T$ is called its consequence.

The consequence operator $(\mbox{-})^{\scriptstyle\bullet}$,
which is defined on specifications,
is a closure operator:
(increasing) $T \subseteq T^{\scriptstyle\bullet}$,
(monotonic)  $T_{1} \subseteq T_{2}$ implies $T_{1}^{\scriptstyle\bullet} \subseteq T_{2}^{\scriptstyle\bullet}$, and
(idempotent) $T^{\scriptstyle\bullet\bullet} = T^{\scriptstyle\bullet}$.
There is an intentional (concept lattice) entailment order between specifications that is implicit in satisfaction:
$T_{1} \leq_{\Sigma} T_{2}$ when $T_{1}^{\scriptstyle\bullet} \supseteq T_{2}^{\scriptstyle\bullet}$;
equivalently,
$T_{1}^{\scriptstyle\bullet} \supseteq T_{2}$.
This is a specialization-generalization order;
$T_{1}$ is more specialized than $T_{2}$, and $T_{2}$ is more generalized than $T_{1}$.
We symbolize this preorder by 
$\mathrmbfit{fbr}(\Sigma)^{\mathrm{op}} = \langle \mathrmbfit{spec}(\Sigma), \leq_{\Sigma} \rangle$.
Intersections and unions define joins and meets.
Its opposite preorder is symbolize by 
$\mathrmbfit{fbr}(\Sigma) = \langle{\mathrmbfit{spec}(\Sigma),\geq_{\Sigma}}\rangle$.
Any specification $T$ is entailment equivalent to its consequence $T \cong T^{\scriptstyle\bullet}$.
A specification $T$ is closed when it is equal to its consequence $T = T^{\scriptstyle\bullet}$.
This paper is concerned with extending the notion of consequence from specifications to information systems.
Although implicit,
we usually include the language in the symbolism,
so that a {\em specification} (presentation) $\mathcal{T}	= \langle \Sigma, T \rangle$ is an indexed notion
consisting of a language $\Sigma$ and a $\Sigma$-specification $T \in \mathrmbfit{spec}(\Sigma)$.$^\text{\ref{homogenization:process}}$

Inverse image preserves closed specifications:
for any language morphism $\sigma : \Sigma_{1} \rightarrow \Sigma_{2}$, 
${(-)}^{\scriptscriptstyle\bullet} \circ \mathrmbfit{sen}(\sigma)^{-1} \circ {(-)}^{\scriptscriptstyle\bullet} 
= {(-)}^{\scriptscriptstyle\bullet} \circ \mathrmbfit{sen}(\sigma)^{-1}$; 
that is, 
 $\mathrmbfit{sen}(\sigma)^{-1}(T_{2}^{\scriptscriptstyle\bullet})^{\scriptscriptstyle\bullet} 
= \mathrmbfit{sen}(\sigma)^{-1}(T_{2}^{\scriptscriptstyle\bullet})$
for any target specification $T_{2} \in \mathrmbfit{spec}(\Sigma_{2})$. 
Direct image commutes with consequence:
for any language morphism $\sigma : \Sigma_{1} \rightarrow \Sigma_{2}$, 
${\wp}\mathrmbfit{sen}(\sigma) \circ {(-)}^{\scriptscriptstyle\bullet}
= {(-)}^{\scriptscriptstyle\bullet} \circ {\wp}\mathrmbfit{sen}(\sigma) \circ {(-)}^{\scriptscriptstyle\bullet}$; 
that is,
${\wp}\mathrmbfit{sen}(\sigma)(T_{1})^{\scriptscriptstyle\bullet}
= {\wp}\mathrmbfit{sen}(\sigma)(T_{1}^{\scriptscriptstyle\bullet})^{\scriptscriptstyle\bullet}$
for any source specification $T_{1} \in \mathrmbfit{spec}(\Sigma_{1})$.
The formal molecules (specifications) 
can be moved along language morphisms.
For any language morphism $\sigma : \Sigma_{1} \rightarrow \Sigma_{2}$,
define the {\em direct flow} operator
$\mathrmbfit{dir}(\sigma) = {\wp}\mathrmbfit{sen}(\sigma)
: \mathrmbfit{spec}(\Sigma_{1}) \rightarrow \mathrmbfit{spec}(\Sigma_{2})$
and the {\em inverse flow} operator
$\mathrmbfit{inv}(\sigma) = 
\mathrmbfit{sen}(\sigma)^{-1}((\mbox{-})^{\scriptscriptstyle\bullet})
: \mathrmbfit{spec}(\Sigma_{2}) \rightarrow \mathrmbfit{spec}(\Sigma_{1})$.
These are adjoint monotonic functions w.r.t. specification order:
$\mathrmbfit{inv}(\sigma)(T_{2}) \leq_{\Sigma_{1}} T_{1} 
\text{ \underline{iff} } 
T_{2} \leq_{\Sigma_{2}} \mathrmbfit{dir}(\sigma)(T_{1})$.\footnote{The paper 
(Tarlecki, et al \cite{tarlecki:burstall:goguen:91})
claims that inverse image can be used without first computing the consequence.} 
We symbolized this adjunction by
$\langle \mathrmbfit{inv}(\sigma), \mathrmbfit{dir}(\sigma) \rangle 
: \mathrmbfit{fbr}(\Sigma_{2}) \rightarrow \mathrmbfit{fbr}(\Sigma_{1})$
between entailment preorders
or by
$\langle \mathrmbfit{dir}(\sigma), \mathrmbfit{inv}(\sigma) \rangle 
: \mathrmbfit{fbr}(\Sigma_{1})^{\mathrm{op}} \rightarrow \mathrmbfit{fbr}(\Sigma_{2})^{\mathrm{op}}$
between opposite preorders.
Hence, there are indexed categories
$\mathrmbfit{dir} : \mathrmbf{Lang} \rightarrow \mathrmbf{Set}$
and
$\mathrmbfit{inv} : \mathrmbf{Lang}^{\mathrm{op}} \rightarrow \mathrmbf{Set}$
for specifications.

A specification morphism
$\sigma : \mathcal{T}_{1} \rightarrow \mathcal{T}_{2}$
is a language morphism $\sigma : \Sigma_{1} \rightarrow \Sigma_{2}$
that preserves entailment:
$T_{1} \vdash_{\Sigma_{1}} s_{1}$
implies 
$T_{2} \vdash_{\Sigma_{2}} \mathrmbfit{sen}(\sigma)(s_{1})$
for any $s_{1} \in \mathrmbfit{sen}(\Sigma_{1})$.
Equivalently,
that maps the source specification to a generalization of the target specification
$\mathrmbfit{dir}(\sigma)(T_{1}) 
\geq_{\Sigma_{2}} T_{2}$;
or
that maps the target specification to a specialization of the source specification
$T_{1} \geq_{\Sigma_{1}} \mathrmbfit{inv}(\sigma)(T_{2})$.
Thus,
the fibered category of specifications $\mathrmbf{Spec}$
is defined in terms of formal information flow.
The fibered category of specifications $\mathrmbf{Spec}$ 
has specifications as objects and specification morphisms as morphisms (Fig.~\ref{logical:system}).
There is an underlying language functor
$\mathrmbfit{lang} : \mathrmbf{Spec} \rightarrow \mathrmbf{Lang}$ from specifications to languages 
$\mathcal{T} = \langle \Sigma, T \rangle \mapsto \Sigma$.

\subsubsection{Structures.}\label{structures}

For a language $\Sigma$,
the conceptual (concept lattice) intent of a $\Sigma$-structure $M \in \mathrmbfit{struc}(\Sigma)$, 
implicit in satisfaction,
is the (closed) specification 
$M^{\Sigma} = \{ s \in \mathrmbfit{sen}(\Sigma) \mid M \models_{\Sigma} s \}$
consisting of all sentences satisfied by the structure. 
There is an intentional (concept lattice) order between $\Sigma$-structures:
$M_{1} \leq_{\Sigma} M_{2}$ when 
$M_{1}^{\Sigma} \leq_{\Sigma} M_{2}^{\Sigma}$ (specification order);
equivalently,
$M_{1}^{\Sigma} \supseteq M_{2}^{\Sigma}$.
An indexed {\em structure} $\mathcal{M}	= \langle \Sigma, M \rangle$ consists of 
a language $\Sigma$ and a $\Sigma$-structure $M$.$^\text{\ref{homogenization:process}}$

An indexed structure morphism
$\sigma 
: \mathcal{M}_{1} = \langle \Sigma_{1}, M_{1} \rangle \rightarrow \langle \Sigma_{2}, M_{2} \rangle = \mathcal{M}_{2}$
is a language morphism $\sigma : \Sigma_{1} \rightarrow \Sigma_{2}$
that preserves satisfaction:
$M_{1} \models_{\Sigma_{1}} s_{1}$ implies $M_{2} \models_{\Sigma_{2}} \mathrmbfit{sen}(\sigma)(s_{1})$
for any $s_{1} \in \mathrmbfit{sen}(\Sigma_{1})$;
that is,
$\mathrmbfit{sen}(\sigma)^{-1}(M_{2}^{\Sigma_{2}}) \leq_{\Sigma_{1}} M_{1}^{\Sigma_{1}}$
meaning
$\sigma : \langle \Sigma_{1}, M_{1}^{\Sigma_{1}} \rangle \rightarrow \langle \Sigma_{2}, M_{2}^{\Sigma_{2}} \rangle$
is a specification morphism.
Equivalently,
(by satisfaction invariance)
$\mathrmbfit{struc}(\sigma)(M_{2})^{\Sigma_{1}} \leq_{\Sigma_{1}} M_{1}^{\Sigma_{1}}$
or (by definiton of structure order)
that maps the target structure to a specialization of the source structure
$\mathrmbfit{struc}(\sigma)(M_{2}) \leq_{\Sigma_{1}} M_{1}$.
The fibered category of structures $\mathrmbf{Struc}$ 
has indexed structures as objects and structure morphisms as morphisms (Fig.~\ref{logical:system}).
There is an underlying language functor 
$\mathrmbfit{lang} : \mathrmbf{Struc} \rightarrow \mathrmbf{Lang}$ from structures to languages 
$\mathcal{M} = \langle \Sigma, M \rangle \mapsto \Sigma$.
Also,
there is a conceptual intent functor 
$\mathrmbfit{int} : \mathrmbf{Struc} \rightarrow \mathrmbf{Spec}$ from structures to specifications, 
where
$\mathrmbfit{int} \circ \mathrmbfit{lang} = \mathrmbfit{lang}$. 

\subsubsection{Logics.}\label{logics}

A populated ontology expressed in terms of a language $\Sigma$
is represented as a (generic) {\em logic} $\mathcal{L}$ 
having two components, 
a structure $M \in \mathrmbfit{struc}(\Sigma)$
and a specification $T \in \mathrmbfit{spec}(\Sigma)$
that share $\Sigma$.\footnote{Using only a single structure in logics is not a restriction.
For any classification $A = \langle X, Y, \models \rangle$
there is a power instance classification ${\wp}A = \langle {\wp}X, Y, \models_{\wp} \rangle$
where $\check{X} \models_{\wp} y$ holds when $x \models y$ for all $x \in \check{X}$.
For any infomorphism $f = \langle \check{f}, \hat{f} \rangle : A_{1} \rightleftharpoons A_{2}$
there is a power instance infomorphism
${\wp}f = \langle {\wp}\check{f}, \hat{f} \rangle : {\wp}A_{1} \rightleftharpoons {\wp}A_{2}$
with direct image instance function.
Combining these constructions defines a power instance functor
${\wp}\mathrmbfit{inst} : \mathrmbf{Cls} \rightarrow \mathrmbf{Cls}$.
Hence, 
for any institution with classification functor
$\mathrmbfit{cls} : \mathrmbf{Lang} \rightarrow \mathrmbf{Cls}$,
there is an associated institution with classification functor
$\mathrmbfit{cls} \circ {\wp}\mathrmbfit{inst} : \mathrmbf{Lang} \rightarrow \mathrmbf{Cls}$.
For this power structure institution
a logic $\mathcal{L} = \langle \Sigma, \mathrmbfit{M}, T \rangle$
consists of 
a collection of structures $\mathrmbfit{M} \subseteq \mathrmbfit{struc}(\Sigma)$ and 
a specification $T \in \mathrmbfit{spec}(\Sigma)$,
where the individual structures in $\mathrmbfit{M}$ may or may not model the specification $T$.
The logic $\mathcal{L}$ is sound when they all model the specification.}
In the institutional approach,
a logic is a molecule of semantics.
Although implicit,
we include the language in the symbolism,
so that a {\em logic} $\mathcal{L}	= \langle \Sigma, M, T \rangle$ is an indexed notion consisting of 
a language $\Sigma$, 
a $\Sigma$-structure $M$ and 
a $\Sigma$-specification $T$.\footnote{\label{homogenization:process}
Languages index structures, specifications and logics.
The homogenization process (Fig.~\ref{logical:system}) (also called the Grothendieck construction), 
moving from indexed categories to fibered categories,
is the process of combining an indexing language $\Sigma$ 
with elements from the indexed category components (fibers) $\mathrmbfit{struc}(\Sigma)$, $\mathrmbfit{spec}(\Sigma)$, etc.}
This notion of logic is a precursor to the local logics defined and used in information flow 
(Barwise and Seligman \cite{barwise:seligman:97}), 
which are represented by the composite logics defined below. 
For any fixed structure $\mathcal{M} = \langle \Sigma, M \rangle$,
the set of all logics $\mathrmbfit{log}(\mathcal{M})$
with that structure is a preordered set under the specification order:
$\langle \Sigma, M, T_{1} \rangle \leq \langle \Sigma, M, T_{2} \rangle$ when $T_{1} \leq T_{2}$.
We denote this preorder by 
$\mathrmbfit{fbr}(\mathcal{M})^{\mathrm{op}} = \langle \mathrmbfit{log}(\mathcal{M}), \leq_{\mathcal{M}} \rangle
\cong \langle \mathrmbfit{spec}(\Sigma), \leq_{\Sigma} \rangle = \mathrmbfit{fbr}(\Sigma)^{\mathrm{op}}$.

A logic morphism 
$\sigma : \mathcal{L}_{1} \rightarrow \mathcal{L}_{2}$
is a language morphism $\sigma : \Sigma_{1} \rightarrow \Sigma_{2}$
that is both a structure morphism 
$\sigma : \langle \Sigma_{1}, M_{1} \rangle \rightarrow \langle \Sigma_{2}, M_{2} \rangle$
and a specification morphism 
$\sigma : \langle \Sigma_{1}, T_{1} \rangle \rightarrow \langle \Sigma_{2}, T_{2} \rangle$. 
The fibered category of logics $\mathrmbf{Log}$ 
has logics as objects and logic morphisms as morphisms (Fig.~\ref{logical:system}).
It is the fibered product of the fibered categories $\mathrmbf{Struc}$ and $\mathrmbf{Spec}$.
There are projective component functors from logics to structures and specifications,
$\mathrmbfit{pr}_0 : \mathrmbf{Log} \rightarrow \mathrmbf{Struc}$ and
$\mathrmbfit{pr}_{1} : \mathrmbf{Log} \rightarrow \mathrmbf{Spec}$,
which satisfy the condition 
$\mathrmbfit{pr}_0 \circ \mathrmbfit{lang} = \mathrmbfit{pr}_{1} \circ \mathrmbfit{lang}$.

The semantic molecules (logics) 
can be moved along structure morphisms.
Define direct and inverse flow of logics 
along structure morphisms
in terms of the specification components.
For any structure morphism 
$\sigma : \mathcal{M}_{1} = \langle \Sigma_{1}, M_{1} \rangle \rightarrow 
          \langle \Sigma_{2}, M_{2} \rangle = \mathcal{M}_{2}$,
define the {\em direct flow} operator
$\mathrmbfit{dir}(\sigma) : \mathrmbfit{log}(\mathcal{M}_{1}) \rightarrow \mathrmbfit{log}(\mathcal{M}_{2})$
by
$\mathrmbfit{dir}(\sigma)(\mathcal{L}_{1})
= \langle \Sigma_{2}, M_{2}, {\wp}\mathrmbfit{sen}(\sigma)(T_{1}) \rangle$
for source logics $\mathcal{L}_{1}	= \langle \Sigma_{1}, M_{1}, T_{1} \rangle$
and the {\em inverse flow} operator
$\mathrmbfit{inv}(\sigma) : \mathrmbfit{log}(\mathcal{M}_{2}) \rightarrow \mathrmbfit{log}(\mathcal{M}_{1})$
by
$\mathrmbfit{inv}(\sigma)(\mathcal{L}_{2})
= \langle \Sigma_{1}, M_{1}, \mathrmbfit{sen}(\sigma)^{-1}({T_{2}}^{\scriptscriptstyle\bullet}) \rangle$
for target logics $\mathcal{L}_{2}	= \langle \Sigma_{2}, M_{2}, T_{2} \rangle$.
These are adjoint monotonic functions w.r.t. logic order:
$\mathrmbfit{inv}(\sigma)(\mathcal{L}_{2}) \leq_{\Sigma_{1}} \mathcal{L}_{1} 
\text{ \underline{iff} } 
\mathcal{L}_{2} \leq_{\Sigma_{1}} \mathrmbfit{dir}(\sigma)(\mathcal{L}_{1})$. 

In general, 
the logics in the institutional approach to information flow 
are neither sound nor complete. 
A logic $\mathcal{L} = \langle \Sigma, M, T \rangle$ is sound 
when the structure models the specification;
equivalently,
$T \vdash s$ implies $M \models s$; 
or $M^{\Sigma} \leq T^{\scriptstyle\bullet}$.
A logic $\mathcal{L}$ is complete 
when every sentence satisfied by the structure is a sentence entailed by the specification,
$M \models s$ implies $T \vdash s$; 
or $T^{\scriptstyle\bullet} \leq M^{\Sigma}$.
A logic $\mathcal{L}$ is sound and complete when structure and specification are ``conceptually'' the same,
$M^{\Sigma} = T^{\scriptstyle\bullet}$,
generating the same concept in the satisfaction concept lattice.
For any structure morphism $\sigma : \mathcal{M}_{1} \rightarrow \mathcal{M}_{2}$,
direct flow preserves soundness:
if a source logic $\mathcal{L}_{1}$ is sound, 
then the direct flow logic $\mathrmbfit{dir}(\sigma)(\mathcal{L}_{1})$ is also sound.
For any structure morphism $\sigma : \mathcal{M}_{1} \rightarrow \mathcal{M}_{2}$,
inverse flow preserves completeness:
if a target logic $\mathcal{L}_{2}$ is complete, 
then the inverse flow logic $\mathrmbfit{inv}(\sigma)(\mathcal{L}_{2})$ is also complete.

\subsubsection{Sound Logics.}\label{sound:logics}

Sound logics form a subcategory of logics
$\mathrmbfit{inc} : \mathrmbf{Snd} \rightarrow \mathrmbf{Log}$
with the same projections (Fig.~\ref{logical:system}).
Associated with any indexed structure $\mathcal{M} = \langle \Sigma, M \rangle$ is a natural logic 
$\mathrmbfit{nat}(\mathcal{M}) = \langle \Sigma, M, M^{\Sigma} \rangle$,
whose specification is the conceptual intent of $\mathcal{M}$.
The natural logic is essentially (up to equivalence) 
the only sound and complete logic over the given language $\Sigma$. 
Any indexed structure morphism
$\sigma : \mathcal{M}_{1} \rightarrow \mathcal{M}_{2}$
induces the natural logic morphism 
$\sigma : \mathrmbfit{nat}(\mathcal{M}_{1}) \rightarrow \mathrmbfit{nat}(\mathcal{M}_{2})$.
Hence,
there is a natural logic functor
$\mathrmbfit{nat} : \mathrmbf{Struc} \rightarrow \mathrmbf{Snd}$.
Structures form a reflective subcategory of sound logics,
since the pair 
$\langle \mathrmbfit{pr}_{0}, \mathrmbfit{nat} \rangle : \mathrmbf{Snd} \rightarrow \mathrmbf{Struc}$
forms an adjunction\footnote{An adjunction (adjoint pair) 
consists of an adjunction of functors; 
that is,
a pair of oppositely-directed functors that satisfy inverse equations up to morphism.
Any ``canonical construction from one species of structure to another'' 
is represented by an adjunction between corresponding categories of the two species (Goguen \cite{goguen:91}).}
with $\mathcal{L} \geq_{\Sigma} \mathrmbfit{nat}(\mathrmbfit{pr}_0(\mathcal{L}))$
and $\mathrmbfit{nat} \circ \mathrmbfit{pr}_{0} = 1_{\mathrmbf{Struc}}$.

Since the identity language morphism $1_{\Sigma} : \Sigma \rightarrow \Sigma$
is a structure morphism $1_{\Sigma} : \langle \Sigma, M_{1} \rangle \rightarrow \langle \Sigma, M_{2} \rangle$
\underline{iff}
$M_{1} \geq_{\Sigma} M_{2}$,
the structure fiber over $\Sigma$ 
w.r.t. $\mathrmbfit{lang} : \mathrmbf{Struc} \rightarrow \mathrmbf{Lang}$
is the opposite of the structure order. 
Since the identity 
\begin{wrapfigure}{r}{155pt}
\setlength{\unitlength}{0.95pt}
\begin{picture}(150,80)(-65,12)
\put(15,102){\makebox(0,0)[r]{\scriptsize{$\top$}}}
\put(26,102){\makebox(0,0)[l]{\scriptsize{$= \langle \Sigma,M,\emptyset \rangle = \top$}}}
\put(15,50){\makebox(0,0)[r]{\scriptsize{$\mathrmbfit{nat}(\mathcal{M})$}}}
\put(26,50){\makebox(0,0)[l]{\scriptsize{$= \langle \Sigma,M,M^{\Sigma} \rangle$}}}
\put(15,-2){\makebox(0,0)[r]{\scriptsize{$\bot$}}}
\put(26,-2){\makebox(0,0)[l]{\scriptsize{$= \langle \Sigma,M,{\wp}\mathrmbfit{sen}(\Sigma) \rangle$}}}
\put(20,72){\makebox(0,0){\tiny{$\emph{sound}$}}}
\put(20,28){\makebox(0,0){\tiny{$\emph{complete}$}}}
\put(-62,50){\makebox(0,0)[l]{\scriptsize{\shortstack{$\mathrmbfit{log}(\mathcal{M})$\\$\cong$\\$\mathrmbfit{spec}(\Sigma)$}}}}
\put(-30,50){\makebox(0,0)[l]{\small{$\left\{ \rule[15pt]{0pt}{30pt} \right.$}}}
\put(20,100){\circle*{3}}
\put(20,50){\circle*{3}}
\put(20,0){\circle*{3}}
\qbezier(-10,50)(-10,90)(20,100)
\qbezier(-10,50)(-10,10)(20,0)
\qbezier(50,50)(50,90)(20,100)
\qbezier(50,50)(50,10)(20,0)
\put(20,50){\line(-2,3){22}}
\put(20,50){\line(2,3){22}}
\qbezier[30](-2,83)(20,90)(42,83)
\qbezier[30](-2,83)(20,76)(42,83)
\put(20,50){\line(-2,-3){22}}
\put(20,50){\line(2,-3){22}}
\qbezier[30](-2,17)(20,24)(42,17)
\qbezier[30](-2,17)(20,10)(42,17)
\end{picture}
\end{wrapfigure}
language morphism $1_{\Sigma} : \Sigma \rightarrow \Sigma$ 
is a specification morphism $1_{\Sigma} : \langle \Sigma, T_{1} \rangle \rightarrow \langle \Sigma, T_{2} \rangle$
\underline{iff} $T_{1} \geq_{\Sigma} T_{2}$,
the specification fiber over $\Sigma$ 
w.r.t. $\mathrmbfit{lang} : \mathrmbf{Spec} \rightarrow \mathrmbf{Lang}$
is the opposite of the specification order
$\mathrmbfit{spec}(\Sigma) = \mathrmbfit{fbr}(\Sigma)^{\mathrm{op}}$. 
For any fixed structure $\mathcal{M} = \langle \Sigma, M \rangle$,
since the identity structure morphism $1_{\mathcal{M}} : \mathcal{M} \rightarrow \mathcal{M}$
is a logic morphism $1_{\mathcal{M}} : \langle \Sigma, M, T_{1} \rangle \rightarrow \langle \Sigma, M, T_{2} \rangle$
\underline{iff} $\langle \Sigma, M, T_{1} \rangle \geq_{\mathcal{M}} \langle \Sigma, M, T_{2} \rangle$
\underline{iff} $T_{1} \geq_{\Sigma} T_{2}$,
the logic fiber over $\mathcal{M}$ 
w.r.t. $\mathrmbfit{pr}_{0} : \mathrmbf{Log} \rightarrow \mathrmbf{Struc}$
is the opposite of the logic order
$\mathrmbfit{log}(\mathcal{M}) = \mathrmbfit{fbr}(\mathcal{M})^{\mathrm{op}} \cong \mathrmbfit{spec}(\Sigma)$. 
There are larger fibers.
For any fixed language $\Sigma$,
the set of all logics 
with that language is a preordered set under the 
structure and specification orders:
$\langle \Sigma, M_{1}, T_{1} \rangle \leq \langle \Sigma, M_{2}, T_{2} \rangle$ 
when $M_{1} \leq_\Sigma M_{2}$ and $T_{1} \leq_\Sigma T_{2}$.
This is the (opposite of the) fiber over $\Sigma$ 
w.r.t. the composite functor $\mathrmbfit{pr}_0 \circ \mathrmbfit{lang} : \mathrmbf{Log} \rightarrow \mathrmbf{Lang}$.

Associated with any logic $\mathcal{L} = \langle \Sigma, M, T \rangle$ is its restriction
$\mathrmbfit{res}(\mathcal{L})	
= \mathcal{L} \vee_{\Sigma} \mathrmbfit{nat}(\mathcal{M})
= \langle \Sigma, M, M^{\Sigma} \cap T^{\scriptstyle\bullet} \rangle$,
which is the conceptual join of the logic with the natural logic of its structure component.
Clearly,
the restriction is a sound logic and
$\mathrmbfit{res}(\mathcal{L}) \geq_{\Sigma} \mathcal{L}$.
There is a restriction functor 
$\mathrmbfit{res} : \mathrmbf{Log} \rightarrow \mathrmbf{Snd}$, 
which maps a logic $\mathcal{L}$ to the sound logic
$\mathrmbfit{res}(\mathcal{L})$ 
and maps a logic morphism $\sigma : \mathcal{L}_{1} \rightarrow \mathcal{L}_{2}$
to the morphism of sound logics
$\mathrmbfit{res}(\sigma) 
= \sigma : \mathrmbfit{res}(\mathcal{L}_{1}) \rightarrow \mathrmbfit{res}(\mathcal{L}_{2})$.
This is well-defined since it just couples the structure morphism condition to the theory morphism condition.
The category of sound logics forms a coreflective subcategory of the category of logics,
since the pair 
$\langle \mathrmbfit{inc}, \mathrmbfit{res} \rangle : \mathrmbf{Log} \rightarrow \mathrmbf{Snd}$
forms an adjunction
with $\mathrmbfit{inc} \circ \mathrmbfit{res} = 1_{\mathrmbf{Snd}}$
and $\mathrmbfit{inc}(\mathrmbfit{res}(\mathcal{L}))	\geq_{\Sigma} \mathcal{L}$
for any logic $\mathcal{L}$.
For any structure $\mathcal{M}$,
restriction and inclusion on fibers are adjoint monotonic functions
$\langle \mathrmbfit{res}_{\mathcal{M}}, \mathrmbfit{inc}_{\mathcal{M}} \rangle
: \mathrmbfit{log}(\mathcal{M}) \rightarrow \mathrmbfit{snd}(\mathcal{M})$,
where
$\mathrmbfit{log}(\mathcal{M}) = \mathrmbfit{fbr}^{\mathrm{op}}(\mathcal{M})$
is the opposite fiber of logics over $\mathcal{M}$ 
and
$\mathrmbfit{snd}(\mathcal{M})$ is the opposite fiber of sound logics.
For any structure morphism $\sigma : \mathcal{M}_{1} \rightarrow \mathcal{M}_{2}$,
the restriction-inclusion adjunction on fibers is compatible with the inverse-direct flow adjunction:
$\langle \mathrmbfit{res}_{\mathcal{M}_{2}}, \mathrmbfit{inc}_{\mathcal{M}_{2}} \rangle
\cdot \langle \mathrmbfit{inv}_{\mathrmbf{Snd}}(\sigma), \mathrmbfit{dir}_{\mathrmbf{Snd}}(\sigma) \rangle
= \langle \mathrmbfit{inv}_{\mathrmbf{Log}}(\sigma), \mathrmbfit{dir}_{\mathrmbf{Log}}(\sigma) \rangle
\cdot \langle \mathrmbfit{res}_{\mathcal{M}_{1}}, \mathrmbfit{inc}_{\mathcal{M}_{1}} \rangle$.

The movement of sound logics is a modification of logic flow.
Direct flow preserves soundness, hence there is no change.
Augment inverse flow by restricting to sound logics (joining with structure-intent).
For any structure morphism 
$\sigma : \mathcal{M}_{1} = \langle \Sigma_{1}, M_{1} \rangle \rightarrow 
          \langle \Sigma_{2}, M_{2} \rangle = \mathcal{M}_{2}$,
define the {\em direct flow} operator
$\mathrmbfit{dir}(\sigma) : \mathrmbfit{snd}(\mathcal{M}_{1}) \rightarrow \mathrmbfit{snd}(\mathcal{M}_{2})$
by
$\mathrmbfit{dir}(\sigma)(\mathcal{L}_{1})
= \langle \Sigma_{2}, M_{2}, {\wp}\mathrmbfit{sen}(\sigma)(T_{1}) \rangle$
for sound source logics $\mathcal{L}_{1}	= \langle \Sigma_{1}, M_{1}, T_{1} \rangle$
and the {\em inverse flow} operator
$\mathrmbfit{inv}(\sigma) : \mathrmbfit{snd}(\mathcal{M}_{2}) \rightarrow \mathrmbfit{snd}(\mathcal{M}_{1})$
by
$\mathrmbfit{inv}(\sigma)(\mathcal{L}_{2})
= \langle \Sigma_{1}, M_{1}, \mathrmbfit{sen}(\sigma)^{-1}({T_{2}}^{\scriptscriptstyle\bullet}) \vee M_{1}^{\Sigma_{1}} \rangle$
for sound target logics $\mathcal{L}_{2}	= \langle \Sigma_{2}, M_{2}, T_{2} \rangle$.
These are adjoint monotonic functions w.r.t. sound logic order:
$\mathrmbfit{inv}(\sigma)(\mathcal{L}_{2}) \leq_{\Sigma_{1}} \mathcal{L}_{1} 
\text{ \underline{iff} } 
\mathcal{L}_{2} \leq_{\Sigma_{1}} \mathrmbfit{dir}(\sigma)(\mathcal{L}_{1})$
for all sound target logics $\mathcal{L}_{2}$ 
and sound source logics $\mathcal{L}_{1}$.
 
A {\em composite logic}, 
the abstract representation of the (local) logics of the theory of information flow 
(Barwise and Seligman \cite{barwise:seligman:97}), 
consists of a base logic and a sound logic sharing the same language and specification, 
where any sentence satisfied by the base logic structure is also satisfied by the sound logic structure. 
Composite logics form a category 
with two projective component functors to both logics and sound logics.
Sound logics are justified as legitimate objects of study, 
since they are the common abstract form for both universal algebras and knowledge bases.
In the approach used in this paper, 
generic logics are useful as first steps (precursors) toward the definition of sound and composite logics.
Composite logics represent the commonsense theories of aritifial intelligence (AI).
They are justified by the following argument for unsound or incomplete theories in Barwise and Seligman \cite{barwise:seligman:97}
``Ordinary reasoning is not logically perfect;
there are logical sins of commission (unsound inferences) and of omission (inferences that are sound but not drawn).
Modeling this, AI has had to cope with logics that are both unsound and incomplete.''

\subsection{Examples}\label{examples}

Examples of logical systems
(Goguen and Burstall \cite{goguen:burstall:92}), 
(Mossakowski, et al \cite{mossakowski:goguen:diaconescu:tarlecki:05})
include:  
first order, 
equational, 
Horn clause, 
intuitionistic, 
modal, 
linear, 
higher-order, 
polymorphic, 
temporal, 
process, 
behavioral, 
coalgebraic and 
object-oriented logics.
In this paper,
we describe four important logical systems:
unsorted equational logic {\ttfamily EQ};
information flow $\mathtt{IF}$,
unsorted first-order logic with equality {\ttfamily FOL},
which extends {\ttfamily EQ} and $\mathtt{IF}$, and
the sketch institution $\mathtt{Sk}$.

In equational logic {\ttfamily EQ},
languages are families of function symbols with arity, 
sentences are equations between terms of function symbols,
structures are abstract algebras (universe, plus operations), 
and satisfaction is equational satisfaction.
First-order logic {\ttfamily FOL}
extends equational logic by adding relation symbols,
so that $\mathtt{EQ}$ is a subsystem of $\mathtt{FOL}$:
sentences are the usual first order sentences (equations, relational expressions, connectives, quantifiers),
structures extend those of unsorted equational logic by adding relations with terms,
and satisfaction is as usual.
In information flow $\mathtt{IF}$, 
languages are sets of type symbols, 
language morphisms are maps of type symbols,
sentences are sequents of type symbols,
structures are classifications,
and satisfaction is sequent satisfaction by instances.
$\mathtt{IF}$ is a subsystem of $\mathtt{FOL}$
when types are regarded as unary relation symbols. 
The sketch institution $\mathtt{Sk}$ is the category-theoretic approach 
to ontological specification (Barr and Wells \cite{barr:wells:99}),
whose special cases include 
multisorted universal algebra, 
the entity relationship data model (Johnson and Rosebrugh \cite{johnson:rosebrugh:07}), and 
topos axiomatizations (foundations).

\section{Channel Theory}\label{sec:channel:theory}

\begin{principle}{System}
Information flow results from regularities in a distributed system.
\emph{(This is the first principle of the theory of information flow, 
as discussed in Barwise and Seligman \cite{barwise:seligman:97}.)}
\end{principle}
This principle motivates the representation of distributed systems by diagrams 
of objects that can incorporate regularities.
We will argue that these objects should be 
specifications in a formal representation or logics in a semantic representation.

\subsection{Information Systems}\label{information:systems}

In general systems theory, 
a system is a collection of interconnected parts,
where the whole may have properties 
that cannot be known from an analysis of the constituent parts in isolation.
In an information system,
the parts represent information resources
and the interconnections represent constraints\footnote{In general,
a constraint is conceptually an interconnection between many parts 
(a special case is an $n$-ary relation $R(A_{1},A_{1},\ldots,A_{n})$).
It can be represented with the connective form of an $n$-ary span. 
An $n$-ary span $(f_{k} : R \rightarrow A_{k} \mid 1 \leq k \leq n)$
in a category consists of 
$n$ morphisms (directed binary constraints) $f_{k}$
with a common source or vertex object $R$ (relational concept) 
connecting $n$ component objects $A_{k}$.
Thus,
we represent a system as a diagram 
consisting of a collection of objects and a collection of binary constraints.
Compare (1) the use of thematic roles (case relations) in conceptual graphs 
and (2) the representation of entity-relationship modeling with sketches 
(Johnson and Rosebrugh \cite{johnson:rosebrugh:07}).}
between the parts.

\subsubsection{Example.}\label{example}

Consider a semantic information system consisting of the logics of four communities 
$\mathcal{L}_{0}, \mathcal{L}_{1}, \mathcal{L}_{2}, \mathcal{L}_{3}$ 
that wish to interact in various ways to share some of their information.
We assume that 
$\mathcal{L}_{0}$ and $\mathcal{L}_{1}$ 
would like to collaborate and share information through a binary span
with vertex (reference or bridging) sublogic $\mathcal{C}$.
Assume the same is true for
$(\mathcal{B}, \mathcal{L}_{0}, \mathcal{L}_{2})$,
$(\mathcal{D}, \mathcal{L}_{1}, \mathcal{L}_{3})$, and
$(\mathcal{E}, \mathcal{L}_{2}, \mathcal{L}_{3})$.
We also assume that
$\mathcal{L}_{0}$, $\mathcal{L}_{1}$ and $\mathcal{L}_{2}$
would like to collaborate and share information through a ternary span
with vertex sublogic $\mathcal{A}$.
Hence,
the total information system consists of nine logics and eleven logic morphisms:
the four community logics mentioned above,
plus five mediating logics 
$\mathcal{A}$, $\mathcal{B}$, $\mathcal{C}$, $\mathcal{D}$ and $\mathcal{E}$
and eleven linking logic morphisms 
$\mathcal{L}_{0}{\,\leftarrow\,}\mathcal{C}{\,\rightarrow\,}\mathcal{L}_{1}$, 
\ldots, $\mathcal{A}{\,\rightarrow\,}\mathcal{L}_{0}, \ldots \,$
making up the spans.
The underlying distributed system has the same shape,
and consists of nine structures and eleven structure morphisms:
four community structures 
$\mathcal{M}_{0} = \mathrmbfit{pr}_{0}(\mathcal{L}_{0})$, \ldots 
underlying the community logics,
plus five  reference structures 
$\mathrmbfit{pr}_{0}(\mathcal{A})$, \ldots,
and eleven linking structure morphisms 
$\mathcal{M}_{0}{\,\leftarrow\,}\mathrmbfit{pr}_{0}(\mathcal{C}){\,\rightarrow\,}\mathcal{M}_{1}, \ldots \,$
used to help aid the collaboration.

\begin{wrapfigure}{r}{110pt}
\setlength{\unitlength}{0.56pt}
\begin{picture}(180,120)(0,30)
\put(-5,120){
\begin{picture}(60,60)(0,0)
\put(31,30.5){\makebox(0,0){\footnotesize{$\mathcal{L}_{0}$}}}
\put(45,44){\makebox(0,0){\scriptsize{$\mathcal{C}_{0}$}}}
\put(17,16){\makebox(0,0){\scriptsize{$\mathcal{B}_{0}$}}}
\put(45,16){\makebox(0,0){\scriptsize{$\mathcal{A}_{0}$}}}
\put(0,60){\line(1,0){60}}
\put(0,0){\line(1,0){60}}
\put(0,0){\line(0,1){60}}
\put(60,0){\line(0,1){60}}
\put(0,30){\line(1,0){20}}
\put(60,30){\line(-1,0){20}}
\put(30,0){\line(0,1){20}}
\put(30,60){\line(0,-1){20}}
\end{picture}}
\put(-5,0){
\begin{picture}(60,60)(0,0)
\put(31,30.5){\makebox(0,0){\footnotesize{$\mathcal{L}_{2}$}}}
\put(17,44){\makebox(0,0){\scriptsize{$\mathcal{B}_{2}$}}}
\put(45,44){\makebox(0,0){\scriptsize{$\mathcal{A}_{2}$}}}
\put(45,16){\makebox(0,0){\scriptsize{$\mathcal{E}_{2}$}}}
\put(0,60){\line(1,0){60}}
\put(0,0){\line(1,0){60}}
\put(0,0){\line(0,1){60}}
\put(60,0){\line(0,1){60}}
\put(0,30){\line(1,0){20}}
\put(60,30){\line(-1,0){20}}
\put(30,0){\line(0,1){20}}
\put(30,60){\line(0,-1){20}}
\end{picture}}
\put(115,120){
\begin{picture}(60,60)(0,0)
\put(31,30.5){\makebox(0,0){\footnotesize{$\mathcal{L}_{1}$}}}
\put(17,44){\makebox(0,0){\scriptsize{$\mathcal{C}_{1}$}}}
\put(17,16){\makebox(0,0){\scriptsize{$\mathcal{A}_{1}$}}}
\put(45,16){\makebox(0,0){\scriptsize{$\mathcal{D}_{1}$}}}
\put(0,60){\line(1,0){60}}
\put(0,0){\line(1,0){60}}
\put(0,0){\line(0,1){60}}
\put(60,0){\line(0,1){60}}
\put(0,30){\line(1,0){20}}
\put(60,30){\line(-1,0){20}}
\put(30,0){\line(0,1){20}}
\put(30,60){\line(0,-1){20}}
\end{picture}}
\put(115,0){
\begin{picture}(60,60)(0,0)
\put(31,30.5){\makebox(0,0){\footnotesize{$\mathcal{L}_{3}$}}}
\put(45,44){\makebox(0,0){\scriptsize{$\mathcal{D}_{3}$}}}
\put(17,16){\makebox(0,0){\scriptsize{$\mathcal{E}_{3}$}}}
\put(0,60){\line(1,0){60}}
\put(0,0){\line(1,0){60}}
\put(0,0){\line(0,1){60}}
\put(60,0){\line(0,1){60}}
\put(0,30){\line(1,0){20}}
\put(60,30){\line(-1,0){20}}
\put(30,0){\line(0,1){20}}
\put(30,60){\line(0,-1){20}}
\end{picture}}
\put(145,75){
\begin{picture}(30,30)(0,0)
\put(0,30){\line(1,0){30}}
\put(0,0){\line(1,0){30}}
\put(0,0){\line(0,1){30}}
\put(30,0){\line(0,1){30}}
\put(15,15){\makebox(0,0){\scriptsize{$\mathcal{D}$}}}
\put(15,30){\vector(0,1){15}}
\put(15,0){\vector(0,-1){15}}
\end{picture}}
\put(70,150){
\begin{picture}(30,30)(0,0)
\put(0,30){\line(1,0){30}}
\put(0,0){\line(1,0){30}}
\put(0,0){\line(0,1){30}}
\put(30,0){\line(0,1){30}}
\put(15,15){\makebox(0,0){\scriptsize{$\mathcal{C}$}}}
\put(0,15){\vector(-1,0){15}}
\put(30,15){\vector(1,0){15}}
\end{picture}}
\put(70,0){
\begin{picture}(30,30)(0,0)
\put(0,30){\line(1,0){30}}
\put(0,0){\line(1,0){30}}
\put(0,0){\line(0,1){30}}
\put(30,0){\line(0,1){30}}
\put(15,15){\makebox(0,0){\scriptsize{$\mathcal{E}$}}}
\put(0,15){\vector(-1,0){15}}
\put(30,15){\vector(1,0){15}}
\end{picture}}
\put(-5,75){
\begin{picture}(30,30)(0,0)
\put(0,30){\line(1,0){30}}
\put(0,0){\line(1,0){30}}
\put(0,0){\line(0,1){30}}
\put(30,0){\line(0,1){30}}
\put(15,15){\makebox(0,0){\scriptsize{$\mathcal{B}$}}}
\put(15,30){\vector(0,1){15}}
\put(15,0){\vector(0,-1){15}}
\end{picture}}
\put(70,75){
\begin{picture}(30,30)(0,0)
\put(0,30){\line(1,0){30}}
\put(0,0){\line(1,0){30}}
\put(0,0){\line(0,1){30}}
\put(30,0){\line(0,1){30}}
\put(15,15){\makebox(0,0){\scriptsize{$\mathcal{A}$}}}
\qbezier(0,0)(-8,-8)(-15,-15)
\put(-15,-15){\vector(-1,-1){0}}
\qbezier(0,30)(-8,38)(-15,45)
\put(-15,45){\vector(-1,1){0}}
\qbezier(30,30)(38,38)(45,45)
\put(45,45){\vector(1,1){0}}
\end{picture}}
\end{picture}
\end{wrapfigure}
These communities might use several ways to collaborate.
Suppose the three communities in the ternary span
$(\mathcal{A}, \mathcal{L}_{0}, \mathcal{L}_{1}, \mathcal{L}_{2})$,
want to combine their axioms on a common underlying structure.
Here the sublogics $\mathcal{A}_{0}$, $\mathcal{A}_{1}$ and $\mathcal{A}_{2}$
share a common underlying structure with logic $\mathcal{A}$,
so that
 $\mathrmbfit{pr}_{0}(\mathcal{A}_{0})
= \mathrmbfit{pr}_{0}(\mathcal{A}_{1}) 
= \mathrmbfit{pr}_{0}(\mathcal{A}_{2}) 
= \mathrmbfit{pr}_{0}(\mathcal{A})$.
As a concrete example, 
suppose these communities are businesses that want to collaborate about safety.
Logic $\mathcal{A}$ might represent government-mandated rules about safety,
whereas sublogics $\mathcal{A}_{0}$, $\mathcal{A}_{1}$ and $\mathcal{A}_{2}$
might represent special rules that the three businesses
$\mathcal{L}_{0}$, $\mathcal{L}_{1}$ and $\mathcal{L}_{2}$
need for some business transactions.
Suppose the pairs of communities 
in the four binary spans above
want to bring their vocabularies into alignment.
As a concrete example, 
suppose the two community logics in the binary span
$(\mathcal{E}, \mathcal{L}_{2}, \mathcal{L}_{3})$
are government agencies that represent health care for citizens
in sublogics $\mathcal{E}_{2}$ and $\mathcal{E}_{3}$,
where 
$\mathcal{L}_{2}$ uses the term ``personnel'' for a citizen in the language underlying sublogic $\mathcal{E}_{2}$,
but 
$\mathcal{L}_{3}$ uses the term ``worker'' in its sublogic $\mathcal{E}_{3}$.
In order to equivalence this terminology,
they use the reference logic $\mathcal{E}$ with ``citizen'' in its vocabulary,
and then map this via morphisms 
$\mathcal{E}_{2}{\,\leftarrow\,}\mathcal{E}{\,\rightarrow\,}\mathcal{E}_{3}\,$
as ``personnel'' $\leftarrowtail$ ``citizen'' $\rightarrowtail$ ``worker''.

\begin{principle}{Structure}
Information flow crucially involves structures of the world.
\emph{(This is the second principle of the theory of information flow, 
as discussed in Barwise and Seligman \cite{barwise:seligman:97}
and generalized from classifications to structures.)}
\end{principle}
By the world we mean the category of structures,
and we papaphrase the quote in the introduction to
``It is structures in the world, that carry information; the information they carry is in the form of sentences''. 
This principle motivates 
the use of structures as the underlying objects for the logics 
that incorporate the regularities in a distributed system and 
the use of structure morphisms as the underlying morphisms for the logic morphisms
that incorporate the information flow of regularities in a distributed system.

\subsubsection{Systems.}\label{systems}

We have discussed flow links for primitive/composite notions 
(specification flow over language morphisms, and logic flow over structure morphisms) above, and
we will discuss flow links for complex notions (distributed and information systems) below.
Here we discuss constraint links.
Information systems have either specifications or logics
as their information resources,
depending on whether they are formal or semantic in nature.
A semantic information system 
can alternatively use sound logics or composite logics instead of generic logics. 
A formal information system only uses specifications.
Just as every specification has an underlying language and every logic has an underlying structure,
so also every information system has an underlying distributed system.
As such,
distributed systems have either languages or structures for their component parts,
depending on whether they are under a formal or semantic information system.
Without loss of generality, we discuss only semantic systems.

A {\em distributed system} (Fig.~\ref{channel:theory})
is a diagram\footnote{Let $\mathrmbf{V}$ be any category within which we will work. 
Of course, 
one would normally choose a category $\mathrmbf{V}$ that has some useful properties. 
We keep that category fixed throughout the discussion and call it the ambient category. 
We regard the objects and morphisms in the ambient category $\mathrmbf{V}$ to be values that we want to index, 
and we focus on a particular part of the ambient category. 
We use a functor into $\mathrmbf{V}$ for this purpose. 
A {\em diagram} is a functor 
$\mathrmbfit{D} : \mathrmbf{I} \rightarrow \mathrmbf{V}$
from an indexing or shape category $\mathrmbf{I}$ 
into the ambient category $\mathrmbf{V}$. 
The objects in the indexing category are called indexing objects and 
the morphisms are called indexing morphisms.}
$\mathcal{M} : \mathrmbf{I} \rightarrow \mathrmbf{Struc}$
within the ambient category of structures and structure morphisms.
As such,
it consists of an indexed family 
$\{ \mathcal{M}_{i} = \langle \Sigma_{i}, M_{i} \rangle \mid i \in |\mathrmbf{I}| \}$
of structures together with an indexed family 
$\{ \mathcal{M}_{e} = \sigma_e : \mathcal{M}_{i} \rightarrow \mathcal{M}_{j} 
\mid (e : i \rightarrow j) \in \mathrmbf{I} \}$
of structure morphisms.
Two distributed systems with the same shape are pointwise ordered $\mathcal{M} \leq \mathcal{M}'$
when the component structures satisfy the same ordering $\mathcal{M}_{i} \leq \mathcal{M}'_{i}$ 
for all $i \in |\mathrmbf{I}|$.
An {\em information system} (Fig.~\ref{channel:theory})
is a diagram\footnote{The representation of systems as diagrams 
allows for systems of systems, and systems of systems of systems, etc.
Just use diagrams of diagrams. 
However, 
the product-exponential adjointness for functors then allows for 
the conflation of a system of systems into just a system with product indexing shape.}
$\mathcal{L} : \mathrmbf{I} \rightarrow \mathrmbf{Log}$ 
within the category of logics.
This consists of an indexed family of logics
$\{ \mathcal{L}_{i} = \langle \Sigma_{i}, M_{i}, T_{i} \rangle : i \in |\mathrmbf{I}| \}$
and an indexed family of logic morphisms
$\{ \mathcal{L}_{e} = \sigma_{e} : \mathcal{L}_{i} \rightarrow \mathcal{L}_{j} 
\mid (e : i \rightarrow j) \in \mathrmbf{I} \}$.
Two logical systems with the same shape are pointwise ordered $\mathcal{L} \leq \mathcal{L}'$
when the component logics satisfy the same ordering 
$\mathcal{L}_{i} \leq \mathcal{L}'_{i}$ for all $i \in \mathrmbf{I}$.
This is only a preliminary ordering,
since it does not represent the influence of one part of the system upon another.
An information system $\mathcal{L}$ 
with $\mathcal{L}_{i} = \langle \Sigma_{i}, M_{i}, T_{i} \rangle$
has an underlying distributed system
$\mathcal{M} = \mathcal{L} \circ \mathrmbfit{pr}_0$
of the same shape with $\mathcal{M}_{i} = \langle \Sigma_{i}, M_{i} \rangle$.
This underlying passage preserves order.
Distributed and information systems were initially defined in the theory of information flow 
(Barwise and Seligman \cite{barwise:seligman:97})
for the special logical system {\ttfamily IF}.
In this paper we have defined distributed and information systems in any logical system.

\subsection{Information Flow}\label{information:flow}

\subsubsection{Link Types.}\label{link:types}

\begin{wrapfigure}{r}{100pt}
\begin{center}
\setlength{\unitlength}{0.9pt}
\begin{picture}(90,10)(-5,25)
\put(0,0){
\begin{picture}(0,0)(0,0)
\put(2,60){\makebox(0,0)[l]{\footnotesize{$\mathcal{M}$}}}
\qbezier(-3,45)(-3,40)(-10,30)
\qbezier(-10,30)(-19,18)(-1,-1)
\qbezier(-1,-1)(10,-10)(22,10)
\qbezier(22,10)(35,40)(12,51)
\qbezier(12,51)(-2,57)(-3,45)
\put(1,43){\makebox(0,0)[l]{\tiny{$\mathcal{M}_{1}$}}}
\put(6,38){\circle*{2}}
\put(3,26){\makebox(0,0)[r]{\tiny{$\sigma$}}}
\put(6,34){\vector(0,-1){20}}
\put(6,10){\circle*{2}}
\put(1,4){\makebox(0,0)[l]{\tiny{$\mathcal{M}_{2}$}}}
\end{picture}}
\put(10,0){
\begin{picture}(0,0)(0,0)
\put(24,43){\makebox(0,0){\tiny{$\sigma_{1}$}}}
\put(0,38){\vector(1,0){48}}
\put(0,10){\vector(1,0){48}}
\put(24,5){\makebox(0,0){\tiny{$\sigma_{2}$}}}
\end{picture}}
\put(56,0){
\begin{picture}(0,0)(0,0)
\put(2,60){\makebox(0,0)[l]{\footnotesize{$\mathcal{M}'$}}}
\put(2,43){\makebox(0,0)[l]{\tiny{$\mathcal{M}'_{1}$}}}
\put(6,38){\circle*{2}}
\put(9,26){\makebox(0,0)[l]{\tiny{$\sigma'$}}}
\put(6,34){\vector(0,-1){20}}
\put(6,10){\circle*{2}}
\put(2,4){\makebox(0,0)[l]{\tiny{$\mathcal{M}'_{2}$}}}
\end{picture}}
\put(49,-6){
\begin{picture}(0,0)(0,0)
\qbezier(0,15)(-5,20)(-5,30)
\qbezier(-5,30)(-5,40)(0,50)
\qbezier(0,50)(15,70)(30,40)
\qbezier(30,40)(37,30)(30,10)
\qbezier(30,10)(20,-5)(0,15)
\end{picture}}
\end{picture}
\end{center}
\end{wrapfigure}

In this paper,
there are two kinds of links: constraint links and flow links.
We think of constraints as being orthogonal to flow and being of a static nature.
Constraints are used in the alignment of systems of information resources.
For example,
let $\mathcal{L}$ be an information system 
with underlying distributed system $\mathcal{M} = \mathcal{L} \circ \mathrmbfit{pr}_0$ and 
let $\mathcal{M}'$ be another distributed system.
A constraint link $\sigma : \mathcal{L}_{1} \rightarrow \mathcal{L}_{2}$
in information system $\mathcal{L}$
from an information resource $\mathcal{L}_{1}$ located at $\mathcal{M}_{1} = \mathrmbfit{pr}_{0}(\mathcal{L}_{1})$
  to an information resource $\mathcal{L}_{2}$ located at $\mathcal{M}_{2} = \mathrmbfit{pr}_{0}(\mathcal{L}_{2})$
represents the alignment of various elements in $\mathcal{L}_{2}$
with certain elements in $\mathcal{L}_{1}$.
Although we think of constraints as being static
in nature,
there is actually a local flow, either direct or inverse, along a constraint link 
in order to compare the information resources at source and target
to check satisfaction of alignment requirements. 
Flow links are used to specify and compute the fusion (Kent \cite{kent:dagstuhl}) and consequence of systems;
for example,
a flow link $\sigma_{1} : \mathcal{M}_{1} \rightarrow \mathcal{M}'_{1}$
connecting a structure $\mathcal{M}_{\scriptstyle 1}$ in system $\mathcal{M}$
        to a structure $\mathcal{M}'_{\scriptstyle 1}$ in system $\mathcal{M}'$,
can denote the flow of information between systems $\mathcal{M}$ and $\mathcal{M}'$,
either directly from $M_{1}$ to $\mathcal{M}'_{1}$
   or inversely from $M'_{1}$ to $\mathcal{M}_{1}$.
Flow interacts with constraints;
for example,
the flow links 
$\sigma_{1} : \mathcal{M}_{1} \rightarrow \mathcal{M}'_{1}$
and
$\sigma_{2} : \mathcal{M}_{2} \rightarrow \mathcal{M}'_{2}$
connecting the $\mathcal{M}$-constraint
$\sigma : \mathcal{M}_{1} \rightarrow \mathcal{M}_{2}$
to the $\mathcal{M}'$-constraint
$\sigma' : \mathcal{M}'_{1} \rightarrow \mathcal{M}'_{2}$
should satisfy ``preservation of constraints'' 
in the sense that
composition of the direct (or inverse) flow along constraint/flow paths is equal,
$\mathrmbfit{dir}(\sigma) \circ \mathrmbfit{dir}(\sigma_{2}) 
= \mathrmbfit{dir}(\sigma_{1}) \circ \mathrmbfit{dir}(\sigma')$.

\begin{principle}{Connection}
It is by virtue of regularities among connections 
that information about some components of a distributed system carries information about other components.
\emph{(This is the third principle of the theory of information flow, 
as discussed in Barwise and Seligman \cite{barwise:seligman:97}.)}
\end{principle}
This principle motivates the use of logics over structures,
which lift specifications over languages,
to represent information flow over covering channels of a distributed system.

\subsubsection{Channels.}\label{channels}

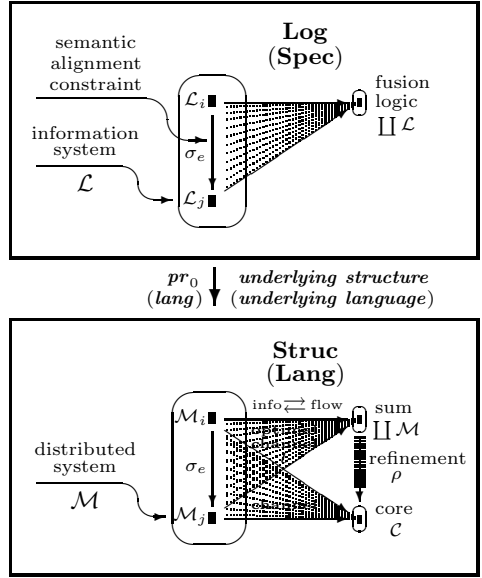
\begin{figure}
\begin{center}
\begin{tabular}{c}
\setlength{\unitlength}{0.8pt}
\begin{picture}(230,280)(0,0)
\put(0,150){
\begin{picture}(220,120)(-14,0)
\thicklines
\put(-14,0){\framebox(220,120){}}
\thinlines
\put(125,106){\makebox(0,0){\footnotesize{$\mathrmbf{Log}$}}}
\put(125,94){\makebox(0,0){\footnotesize{$(\mathrmbf{Spec})$}}}
\put(75,76){\oval(20,20)[tl]}
\put(87,76){\oval(20,20)[tr]}
\put(75,24){\oval(20,20)[bl]}
\put(87,24){\oval(20,20)[br]}
\put(75,86){\line(1,0){12}}
\put(75,14){\line(1,0){12}}
\put(65,24){\line(0,1){52}}
\put(97,24){\line(0,1){52}}
\put(79,71){\rule{3pt}{4pt}}
\put(78.5,74){\makebox(0,0)[r]{\scriptsize{$\mathcal{L}_{i}$}}}
\put(79,24){\rule{3pt}{4pt}}
\put(78.5,27){\makebox(0,0)[r]{\scriptsize{$\mathcal{L}_{j}$}}}
\put(81,67){\vector(0,-1){35}}
\put(78.5,48){\makebox(0,0)[r]{\scriptsize{$\sigma_{e}$}}}
\put(87,73){\vector(1,0){58}}
\qbezier[40](88,71)(119,72)(150,73)
\qbezier[40](88,68)(119,70.5)(150,73)
\qbezier[40](88,65)(119,69)(150,73)
\qbezier[40](88,61)(119,67)(150,73)
\qbezier[40](88,57)(119,65)(150,73)
\qbezier[40](88,53)(119,63)(150,73)
\qbezier[40](88,49)(119,61)(150,73)
\qbezier[40](88,45)(119,59)(150,73)
\qbezier[40](88,41)(119,57)(150,73)
\qbezier[40](88,37)(119,55)(150,73)
\qbezier[40](88,33)(119,53)(150,73)
\put(87,31){\vector(3,2){58}}
\put(150.6,73){\oval(6,12)}
\put(149.6,71){\rule{1.5pt}{3pt}}
\put(158,83){\makebox(0,0)[l]{\scriptsize{$\mathrm{fusion}$}}}
\put(158,73){\makebox(0,0)[l]{\scriptsize{$\mathrm{logic}$}}}
\put(168,62){\makebox(0,0){\scriptsize{$\coprod\mathcal{L}$}}}
\put(46,55){
\begin{picture}(30,20)(0,0)
\put(-24,47){\makebox(0,0){\scriptsize{$\mathrm{semantic}$}}}
\put(-24,36){\makebox(0,0){\scriptsize{$\mathrm{alignment}$}}}
\put(-24,27){\makebox(0,0){\scriptsize{$\mathrm{constraint}$}}}
\put(0,20){\line(-1,0){52}}
\put(0,10){\oval(20,20)[tr]}
\put(20,10){\oval(20,20)[bl]}
\put(20,0){\vector(1,0){8}}
\end{picture}}
\put(30,25){
\begin{picture}(30,20)(0,0)
\put(-13,34){\makebox(0,0){\scriptsize{$\mathrm{information}$}}}
\put(-13,25){\makebox(0,0){\scriptsize{$\mathrm{system}$}}}
\put(4,18){\line(-1,0){40}}
\put(4,10){\oval(16,16)[tr]}
\put(20,10){\oval(16,16)[bl]}
\put(20,2){\vector(1,0){8}}
\put(-13,10){\makebox(0,0){\footnotesize{$\mathcal{L}$}}}
\end{picture}}
\end{picture}}
\thicklines
\put(100,145){\vector(0,-1){18}}
\thinlines
\put(95,140){\makebox(0,0)[r]{\scriptsize{$$}}}
\put(111,140){\makebox(0,0)[l]{\scriptsize{$\mathrmbfit{underlying}$ $\mathrmbfit{structure}$}}}
\put(93,139){\makebox(0,0)[r]{\scriptsize{$\mathrmbfit{pr}_{0}$}}}
\put(107,130){\makebox(0,0)[l]{\scriptsize{$(\mathrmbfit{underlying}$ $\mathrmbfit{language})$}}}
\put(94.5,130){\makebox(0,0)[r]{\scriptsize{$(\mathrmbfit{lang})$}}}
\put(0,0){
\begin{picture}(220,120)(-14,0)
\thicklines
\put(-14,0){\framebox(220,120){}}
\thinlines
\put(125,106){\makebox(0,0){\footnotesize{$\mathrmbf{Struc}$}}}
\put(125,94){\makebox(0,0){\footnotesize{$(\mathrmbf{Lang})$}}}
\put(114,80){\makebox(0,0)[r]{\tiny{$\mathrm{info}$}}}
\put(127.5,80){\makebox(0,0)[l]{\tiny{$\mathrm{flow}$}}}
\put(120,79){\makebox(0,0){\footnotesize{${\rightleftarrows}$}}}
\put(72,76){\oval(20,20)[tl]}
\put(87,76){\oval(20,20)[tr]}
\put(72,24){\oval(20,20)[bl]}
\put(87,24){\oval(20,20)[br]}
\put(72,86){\line(1,0){15}}
\put(72,14){\line(1,0){15}}
\put(62,24){\line(0,1){52}}
\put(97,24){\line(0,1){52}}
\put(79,71){\rule{3pt}{4pt}}
\put(78.75,74){\makebox(0,0)[r]{\scriptsize{$\mathcal{M}_{i}$}}}
\put(79,24){\rule{3pt}{4pt}}
\put(78.75,27){\makebox(0,0)[r]{\scriptsize{$\mathcal{M}_{j}$}}}
\put(81,67){\vector(0,-1){35}}
\put(78.75,50){\makebox(0,0)[r]{\scriptsize{$\sigma_{e}$}}}
\put(87,73){\vector(1,0){58}}
\qbezier[40](88,71)(119,72)(150,73)
\qbezier[40](88,68)(119,70.5)(150,73)
\qbezier[40](88,65)(119,69)(150,73)
\qbezier[40](88,61)(119,67)(150,73)
\qbezier[40](88,57)(119,65)(150,73)
\qbezier[40](88,53)(119,63)(150,73)
\qbezier[40](88,49)(119,61)(150,73)
\qbezier[40](88,45)(119,59)(150,73)
\qbezier[40](88,41)(119,57)(150,73)
\qbezier[40](88,37)(119,55)(150,73)
\qbezier[40](88,33)(119,53)(150,73)
\put(87,31){\vector(3,2){58}}
\put(150.6,73){\oval(6,12)}
\put(149.6,71){\rule{1.5pt}{3pt}}
\put(158,78){\makebox(0,0)[l]{\scriptsize{$\mathrm{sum}$}}}
\put(168,68){\makebox(0,0){\scriptsize{$\coprod\mathcal{M}$}}}
\put(87,68){\vector(3,-2){58}}
\qbezier[40](88,64)(119,45)(150,26)
\qbezier[40](88,62)(119,44)(150,26)
\qbezier[40](88,62)(119,44)(150,26)
\qbezier[40](88,58)(119,42)(150,26)
\qbezier[40](88,54)(119,40)(150,26)
\qbezier[40](88,50)(119,38)(150,26)
\qbezier[40](88,46)(119,36)(150,26)
\qbezier[40](88,42)(119,34)(150,26)
\qbezier[40](88,38)(119,32)(150,26)
\qbezier[40](88,34)(119,30)(150,26)
\qbezier[40](88,30.5)(119,28.5)(150,26.5)
\qbezier[40](88,28)(119,27)(150,26)
\put(87,26){\vector(1,0){58}}
\put(150.6,26){\oval(6,12)}
\put(149.6,24){\rule{1.5pt}{3pt}}
\put(158,30){\makebox(0,0)[l]{\scriptsize{$\mathrm{core}$}}}
\put(168,21){\makebox(0,0){\scriptsize{$\mathcal{C}$}}}
\put(150.7,65){\vector(0,-1){31}}
\qbezier[20](148.6,64.5)(148.6,48)(148.6,41)
\qbezier[20](149.6,64.5)(149.6,48)(149.6,41)
\qbezier[20](151.8,64.5)(151.8,48)(151.8,41)
\qbezier[20](152.8,64.5)(152.8,48)(152.8,41)
\put(155,55){\makebox(0,0)[l]{\scriptsize{$\mathrm{refinement}$}}}
\put(168,46){\makebox(0,0){\scriptsize{$\rho$}}}
\put(30,25){
\begin{picture}(30,20)(0,0)
\put(-13,34){\makebox(0,0){\scriptsize{$\mathrm{distributed}$}}}
\put(-13,25){\makebox(0,0){\scriptsize{$\mathrm{system}$}}}
\put(4,18){\line(-1,0){40}}
\put(4,10){\oval(16,16)[tr]}
\put(20,10){\oval(16,16)[bl]}
\put(20,2){\vector(1,0){5}}
\put(-13,10){\makebox(0,0){\footnotesize{$\mathcal{M}$}}}
\end{picture}}
\put(115,68){\makebox(0,0){\tiny{$\mathrmbf{optimal}$}}}
\put(115,62){\makebox(0,0){\tiny{$\mathrmbf{channel}$}}}
\put(115,55){\makebox(0,0){\footnotesize{$\iota$}}}
\put(115,33){\makebox(0,0){\tiny{$\mathrmbf{channel}$}}}
\end{picture}}
\end{picture}
\end{tabular}
\end{center}
\caption{Channel Theory}
\label{channel:theory}
\end{figure}

For any distributed system $\mathcal{M} : \mathrmbf{I} \rightarrow \mathrmbf{Struc}$,
we think of the component structures $\mathcal{M}_{i}$ as being parts of the system.
We would like to represent the whole system as a structure,
where we might use different structures for different purposes.
The theory of part-whole relations is called mereology.
It studies how parts are related to wholes, 
and how parts are related to other parts within a whole.
In a distributed system, 
the part to part relationships are modeled by the structure morphisms 
$\mathcal{M}_{e} = \sigma_{e} : \mathcal{M}_{i} \rightarrow \mathcal{M}_{j}$ 
indexed by $e : i \rightarrow j$.
We can model the whole as a structure $\mathcal{C}$ 
and model the part-whole relationship 
between some part $\mathcal{M}_{i}$ indexed by $i \in |\mathrmbf{I}|$ and the whole
with a structure morphism $\gamma_{i} : \mathcal{M}_{i} \rightarrow \mathcal{C}$. 
An {\em information channel} 
$\langle \gamma :  \mathcal{M} \Rightarrow \Delta(\mathcal{C}), \mathcal{C} \rangle$
(Fig.~\ref{channel:theory})
(called a corelation by Goguen \cite{goguen:06})
consists of an indexed family
$\{ \gamma_{i} : \mathcal{M}_{i} \rightarrow \mathcal{C} \mid i \in |\mathrmbf{I}| \}$
of structure morphisms with a common target structure $\mathcal{C}$
called the core of the channel\footnote{The notation
$\Delta(\mbox{-})$ denotes the constant operator,
which maps objects to diagrams.
For any structure $\mathcal{A}$, 
the distributed system $\Delta(\mathcal{A}) : \mathrmbf{I} \rightarrow \mathrmbf{Struc}$ is (constantly) 
the structure 
${\Delta(\mathcal{A})}_{i} = \mathcal{A}$ for each index $i \in |\mathrmbf{I}|$ and 
the identity 
${\Delta(\mathcal{A})}_{e} = 1_{\mathcal{A}} : \mathcal{A} \rightarrow \mathcal{A}$ 
for $(e : i \rightarrow j) \in \mathrmbf{I}$.}.
A channel $\langle \gamma, \mathcal{C} \rangle$
{\em covers} a distributed system $\mathcal{M} : \mathrmbf{I} \rightarrow \mathrmbf{Struc}$
when the part-whole relationships respect the system constraints 
(are consistent with the part-part relationships):
$\gamma_{i} = \sigma_{e} \cdot \gamma_{j}$ for each indexing morphism $e : i \rightarrow j$ in $\mathrmbf{I}$.
Covering channels respect the intraconnectivity of the system.
For any two covering channels $\langle \gamma', \mathcal{C}' \rangle$ and $\langle \gamma, \mathcal{C} \rangle$
over the same distributed system $\mathcal{M}$,
a {\em refinement} $\rho : \langle \gamma', \mathcal{C}' \rangle \rightarrow \langle \gamma, \mathcal{C} \rangle$
is a constraint (structure morphism) between cores $\rho : \mathcal{C}' \rightarrow \mathcal{C}$
that respects the part-whole relationships of the two channels:
$\gamma'_{i} \cdot \rho = \gamma_{i}$ for $i \in |\mathrmbf{I}|$.
In such a situation, we say 
the channel $\langle \gamma', \mathcal{C}' \rangle$ 
is a refinement of the channel $\langle \gamma, \mathcal{C} \rangle$.
A channel $\langle \iota, \coprod\mathcal{M} \rangle$ 
is a {\em minimal cover} or {\em optimal(ly refined) channel} of a distributed system $\mathcal{M}$
(Fig.~\ref{channel:theory})
when it covers $\mathcal{M}$ and for any other covering channel $\langle \gamma, \mathcal{C} \rangle$
there is a unique refinement $[\gamma, \mathcal{C}] : \coprod\mathcal{M} \rightarrow \mathcal{C}$
from $\langle \iota, \coprod\mathcal{M} \rangle$ to $\langle \gamma, \mathcal{C} \rangle$.
Any two minimal covers are isomorphic\footnote{In category theory,
a covering channel $\langle \gamma :  \mathcal{M} \Rightarrow \Delta(\mathcal{C}), \mathcal{C} \rangle$
is called a cocone over $\mathcal{M}$, and a minimal cover is called a colimiting cocone over $\mathcal{M}$.}.

\begin{principle}{Channel}
The regularities of a given distributed system 
are relative to its analysis in terms of information channels.
\emph{(This is the fourth principle of the theory of information flow, 
as discussed in Barwise and Seligman \cite{barwise:seligman:97}.)}
\end{principle}
The core of a channel connects the parts of a distributed system,
reflecting the constraints when it covers the system.
More refined means closer connections.

\subsubsection{System Consequence.}\label{system:consequence}

Without loss of generality,
we discuss only the semantic version of system consequence. 
The fibered category $\mathrmbf{Log}$ is cocomplete and
its projection functor $\mathrmbfit{pr}_{0} : \mathrmbf{Log} \rightarrow \mathrmbf{Struc}$ is cocontinuous,
since the fibers $\mathrmbfit{fbr}(\mathcal{M})$ are complete preorders
for all indexing structures $\mathcal{M}$,
direct and inverse flow are adjoint monotonic functions
$\langle \mathrmbfit{dir}(\sigma), \mathrmbfit{inv}(\sigma) \rangle
: \mathrmbfit{fbr}(\mathcal{M}_{1}) \rightarrow \mathrmbfit{fbr}(\mathcal{M}_{2})$
for all indexing structure morphisms $\sigma : \mathcal{M}_{1} \rightarrow \mathcal{M}_{2}$,
and $\mathrmbf{Struc}$ is cocomplete (minimal covers exist for any distributed system).
Then,
information flow can be used to compute colmits in $\mathrmbf{Log}$ 
and to define system consequence. 
This holds also for sound and composite logics.
It holds in the formal version and we can define system consequence for formal information systems,
since comparable properties hold for the fibers $\mathrmbfit{fbr}(\Sigma)$ and the category $\mathrmbf{Lang}$.
This holds for the logical systems $\mathtt{IF}$, $\mathtt{EQ}$ and $\mathtt{FOL}$,
and the special cases of $\mathtt{Sk}$ mentioned above.
It is based upon the colimit theorem 
(Tarlecki, et al \cite{tarlecki:burstall:goguen:91}), 
a general criterion for when such colimits of specifications and logics actually exist.
Let $\mathcal{L} : \mathrmbf{I} \rightarrow \mathrmbf{Log}$ be an information system
with underlying distributed system 
$\mathcal{M} = \mathcal{L} \circ \mathrmbfit{pr}_{0} : \mathrmbf{I} \rightarrow \mathrmbf{Struc}$ 
and optimal (minimal covering) channel $\langle \iota, \coprod\mathcal{M} \rangle$.
The optimal core $\coprod\mathcal{M}$ is called the sum of the distributed system 
$\mathcal{M}$,
and the optimal channel components (structure morphisms)
$\iota_{i} : \mathcal{M}_{i} \rightarrow \coprod\mathcal{M}$
for $i \in |\mathrmbf{I}|$
are flow links.
Fusion and consequence
represent the component ``logics of the system and the way they fit together'' \cite{barwise:seligman:97}.

The {\em fusion} (unification) $\coprod\mathcal{L}$ of the information system $\mathcal{L}$
represents the whole system in a centralized fashion.
This is called fusion in Kent \cite{kent:dagstuhl} and theory blending in Goguen \cite{goguen:06}. 
The fusion logic is defined as {\em direct system flow} (unification).
Direct system flow has two steps:
(i) direct logic flow of the component parts $\{ \mathcal{L}_{i} \mid i \in |\mathrmbf{I}| \}$ 
of the information system 
along the optimal channel over the underlying distributed system
to a centralized location 
(the lattice $\mathrmbfit{fbr}(\coprod\mathcal{M})$ at the optimal channel core $\coprod\mathcal{M}$), and 
(ii) lattice meet combining the contributions of the parts into a whole:
$\coprod\mathcal{L} \doteq \bigwedge \{ \mathrmbfit{dir}(\iota_{i})(\mathcal{L}_{i}) \mid i \in |\mathrmbf{I}| \}$.

The {\em consequence} $\mathcal{L}^{\scriptscriptstyle\blacklozenge}$
of the information system $\mathcal{L}$
represents the whole system in a distributed fashion.
This is an information system,
defined as {\em inverse system flow} (projective distribution).
Inverse system flow has two steps:
(i) consequence of the fusion logic, and
(ii) inverse logic flow of this consequence back along the same optimal channel,
transfering the implications (theorems) of the whole system (the fusion logic)
to the distributed locations $\mathcal{M}_{i}$ of the component parts:
$\mathcal{L}^{\scriptscriptstyle\blacklozenge}
\doteq \{ \mathrmbfit{inv}(\iota_{i})(\coprod\mathcal{L}) \mid i \in |\mathrmbf{I}| \} 
: \mathrmbf{I} \rightarrow \mathrmbf{Log}$.
The consequence operator $(\mbox{-})^{\scriptscriptstyle\blacklozenge}$,
which is defined on information systems,
is a closure operator:
(increasing) $\mathcal{L} \geq \mathcal{L}^{\scriptscriptstyle\blacklozenge}$,
(monotonic)  $\mathcal{L}_{1} \geq \mathcal{L}_{2}$ implies $\mathcal{L}_{1}^{\scriptscriptstyle\blacklozenge} \geq \mathcal{L}_{2}^{\scriptscriptstyle\blacklozenge}$ and
(idempotent) $\mathcal{L}^{\scriptscriptstyle\blacklozenge\blacklozenge} = \mathcal{L}^{\scriptscriptstyle\blacklozenge}$.~\footnote{By allowing system shape to vary,
channels can be generalized to (co)morphisms of distributed systems.
Then a notion of relative fusion (direct system flow) can be defined in terms of left Kan extension,
and a notion of relative system consequence can be defined as 
the composition of direct followed by inverse system flow.}

This is a true abstract system consequence operator, and
an improvement over the ``distributed logic'' operator 
in Lecture (Chapter) 15 of Barwise and Seligman \cite{barwise:seligman:97}
for three reasons:
it maps an information system to another information system,
recognizing the existence of constraint links between the indexed components of the system consequence; 
it recognizes the fact that system consequence is a closure operator, 
satisfying the monotonicity, increasing and idempotency laws; and
it is a true generalization from the specific logical system {\ttfamily IF} to an arbitrary logical system.
Fig.~\ref{channel:theory} provides a graphic representation for the calculation of system consequence:
in the $\mathrmbf{Log}$ category 
information systems are illustrated as ovals, 
ontologies represented by logics are illustrated as nodes within ovals, and 
alignment constraints between ontologies are illustrated as edges between nodes; and 
in the $\mathrmbf{Struc}$ category 
distributed systems are illustrated as ovals, 
structures are illustrated as nodes within ovals, and 
channels are illustrated as triangular shapes. 

Any information system
$\mathcal{L} : \mathrmbf{I} \rightarrow \mathrmbf{Log}$
restricts as the sound information system
$\mathrmbfit{res}\,\mathcal{L} = \mathcal{L} \circ \mathrmbfit{res} : \mathrmbf{I} \rightarrow \mathrmbf{Snd}$,
where each component logic $\mathcal{L}_{i}$ restricts as 
the sound component logic $\mathrmbfit{res}_{\Sigma_{i}}(\mathcal{L}_{i})$ for each $i \in \mathrmbf{I}$.
Any sound information system
$\mathcal{L} : \mathrmbf{I} \rightarrow \mathrmbf{Snd}$
is included as the (generic) information system
$\mathrmbfit{inc}\,\mathcal{L} = \mathcal{L} \circ \mathrmbfit{inc} : \mathrmbf{I} \rightarrow \mathrmbf{Log}$,
where each sound component logic $\mathcal{L}_{i}$ is included as 
the (generic) logic $\mathrmbfit{inc}_{\Sigma_{i}}(\mathcal{L}_{i})$ for each $i \in \mathrmbf{I}$.
Two questions arise.
(1) How is the system consequence $\mathcal{L}^{\scriptscriptstyle\blacklozenge}$ 
of a sound information system $\mathcal{L}$ 
related to 
the system consequence $(\mathrmbfit{inc}\,\mathcal{L})^{\scriptscriptstyle\blacklozenge}$ 
of its inclusion $\mathrmbfit{inc}\,\mathcal{L}$?
The direct system flow along a channel of a sound information system is sound,
and hence the system consequence of a sound information system $\mathcal{L}$ 
is the restriction of the system consequence of its inclusion:
$\mathcal{L}^{\scriptscriptstyle\blacklozenge}
= \mathrmbfit{res}\,(\mathrmbfit{inc}\,\mathcal{L})^{\scriptscriptstyle\blacklozenge}$.
(2) How is the system consequence $\mathcal{L}^{\scriptscriptstyle\blacklozenge}$ 
of an information system $\mathcal{L}$ 
related to 
the system consequence $(\mathrmbfit{res}\,\mathcal{L})^{\scriptscriptstyle\blacklozenge}$
of its sound restriction $\mathrmbfit{res}\,\mathcal{L}$?
In general,
since the sound restriction of the fusion logic of an information system
is more specialized than 
the fusion logic of its sound restriction 
$\mathrmbfit{res}(\coprod\mathcal{L}) \leq \coprod(\mathrmbfit{res}\,\mathcal{L})$,
the sound restriction of the system consequence of an information system $\mathcal{L}$ 
is more specialized than 
the system consequence of its sound restriction 
$\mathrmbfit{res}(\mathcal{L}^{\scriptscriptstyle\blacklozenge}) \leq 
(\mathrmbfit{res}\,\mathcal{L})^{\scriptscriptstyle\blacklozenge}$.
An authentic example showing strict inequality,
would demonstrate that restriction before fusion loses information;
thus providing strong justification for the use of unsound/incomplete logics.

The pointwise entailment order $\leq$ is only a preliminary order,
since it does not incorporate interactions between system component parts.
Just as system consequence  ${(-)}^{\scriptscriptstyle\blacklozenge}$ is analogous to specification consequence ${(-)}^{\scriptscriptstyle\bullet}$,
we think of $\leq$ as analogous to $\supseteq$,
reverse subset order for specifications.
Extending this analogy,
system entailment order 
$\mathcal{L}_{1} \preceq \mathcal{L}_{2}$
for any two $I$-shaped information systems
$\mathcal{L}_{1},\mathcal{L}_{2} : \mathrmbf{I} \rightarrow \mathrmbf{Log}$
is defined by 
$\mathcal{L}_{1}^{\scriptscriptstyle\blacklozenge} \leq \mathcal{L}_{2}^{\scriptscriptstyle\blacklozenge}$;
equivalently,
$\mathcal{L}_{1}^{\scriptscriptstyle\blacklozenge} \leq \mathcal{L}_{2}$.
Pointwise order is stronger than system entailment order: 
$\mathcal{L}_{1} \leq \mathcal{L}_{2}$ implies $\mathcal{L}_{1} \preceq \mathcal{L}_{2}$.
System entailment is a preorder:
(reflexive)
$\mathcal{L} \preceq \mathcal{L}$ and
(transitive) 
if $\mathcal{L}_{1} \preceq \mathcal{L}_{2}$  and $\mathcal{L}_{2} \preceq \mathcal{L}_{3}$, 
then $\mathcal{L}_{1} \preceq \mathcal{L}_{3}$.
This is a specialization-generalization order;
$\mathcal{L}_{1}$ is more specialized than $\mathcal{L}_{2}$, and 
$\mathcal{L}_{2}$ is more generalized than $\mathcal{L}_{1}$.
Any information system $\mathcal{L}$ is entailment equivalent to its consequence 
$\mathcal{L} \cong \mathcal{L}^{\scriptscriptstyle\blacklozenge}$.
An information system $\mathcal{L}$ is closed when it is equal to its consequence 
$\mathcal{L} = \mathcal{L}^{\scriptscriptstyle\blacklozenge}$.

A specific example of system consequence
in $\mathtt{IF}$, the logical system of information flow,
is the file copying example
in Lecture (Chapter) 15 of Barwise and Seligman \cite{barwise:seligman:97},
which involves file properties such as content, time stamp, protection, etc.
Two general examples of system consequence occur in any logical system
$\mathcal{L} : \mathrmbf{I} \rightarrow \mathrmbf{Log}$.
The first is a system with a discrete shape $\mathrmbf{I} = I$,
which is essentially an indexing set.
Then the system consequence is the pointwise consequence
$\mathcal{L}^{\scriptscriptstyle\blacklozenge} = \{ \mathcal{L}_{i}^{\scriptstyle\bullet} \mid i \in I \}$.
The second is a system with any shape $\mathrmbf{I}$, but constant underlying distributed system 
$\Delta(\mathcal{M}) : \mathrmbf{I} \rightarrow \mathrmbf{Struc}$ 
for some single structure $\mathcal{M}$.
Then 
the minimal cover is identity, 
direct system flow is the meet operation (specification union),
inverse system flow is specification consequence, and
system consequence is the constant information system
$\Delta({\left( \bigwedge \mathcal{L}_{i} \right)}^{\scriptstyle\bullet}) : \mathrmbf{I} \rightarrow \mathrmbf{Log}$.
A (formal) concrete example
of this system consequence occurs 
in $\mathtt{FOL}$, the logical system of first order logic,
where the information system 
$\mathcal{T} 
= ( T_{\circ} \stackrel{1}{\leftarrow} T_{\mbox{\tiny{1}}} \stackrel{1}{\rightarrow} T_{\propto} )$
with span shape $\mathrmbf{I} = {\cdot\leftarrow\cdot\rightarrow\cdot}$
consists of the three specifications for 
reflexive relations $T_{\mbox{\tiny{1}}}$,
preorders (reflexive-transitive relations) $T_{\circ}$ and
reflexive-symmetric relations $T_{\propto}$,
with $T_{\mbox{\scriptsize{1}}}$ a subspecification of the other two.
The underlying language for all three specifications is a single binary relation symbol.
The fusion specification is $\mathcal{T}^{\scriptscriptstyle\blacklozenge} = T_{\equiv}$
the (closed) specification for equivalence relations (reflexive-symmetric-transitive relations),
and the system consequence is the constant information system
$\Delta(T_{\equiv}) : \mathrmbf{I} \rightarrow \mathrmbf{Spec}$.

\section{Conclusion}

This paper has discussed system consequence,
one step in the program to combine and extend the theories of institutions and information flow.
The institutional approach was first formulated by Goguen and Burstall \cite{goguen:burstall:92}.
Revealing its importance,
many people have either independently discovered or implicitly used the institutional approach.
The theory of information flow is one example of this.
The Information Flow Framework \cite{iff} 
has followed many of the ideas of information flow, 
and hence has implicitly followed the institutional approach.
The paper Goguen \cite{goguen:06} is an excellent survey of 
the institutional approach to information integration.
However,
the current paper, 
in contrast to Goguen \cite{goguen:06},
believes that information flow follows, and is in great accord with, the institutional approach.
An indication of this accord is revealed by 
the ``Interpretations in First-Order Logic'' example in Barwise and Seligman \cite{barwise:seligman:97}.
We do not believe that institutions are more abstract than information flow,
but that the theory of institutions has not been fully applied in order to generalize the theory of information flow, 
and that the theory of information flow has not been fully applied in order to extend the theory of institutions.
That is the goal of this paper.

This paper has combined two approaches to universal logic, 
the theories of information flow and institutions, 
and has applied them to information systems.
We have given dual descriptions (heterogeneous and homogeneous) of logical systems
and have demonstrated how important concepts in the theory of information flow
can be defined within any logical system:
distributed/information systems, channels, information flow, 
all leading up to system consequence.
Hence,
these concepts are independent of the particular logical system in which one works.
Thus,
the theory of institutions generalizes the theory of information flow, and
the theory of information flow extends the theory of institutions.
A central problem of distributed logic is to understand how one part of a distributed system affects another part.
This paper has solved this problem in the general case of any logical system.
The solution is expressed in terms of system consequence.

\end{document}